\documentclass[11pt]{article}

\usepackage{amsmath,amsfonts,amssymb,amsthm}
\usepackage{mathrsfs}
\usepackage{color}

\definecolor{labelkey}{gray}{.8}
\definecolor{refkey}{gray}{.8}

\definecolor{darkred}{rgb}{0.9,0.1,0.1}
\definecolor{darkgreen}{rgb}{0,0.5,0}

\newtheorem{theorem}{Theorem}[section]
\newtheorem{lemma}[theorem]{Lemma}

\newtheorem{proposition}[theorem]{Proposition}

\theoremstyle{remark}
\newtheorem{remark}[theorem]{Remark}

\numberwithin{equation}{section}
\newcommand{\R}{\mathbb{R}}

\newcommand{\E}{\mathbb{E}}
\newcommand{\F}{\mathcal{F}}
\newcommand{\FF}{\mathscr{F}}

\newcommand{\G}{\mathcal{G}}
\newcommand{\I}{\mathcal{I}}

\newcommand{\V}{\mathcal{V}}

\newcommand{\U}{\mathcal{U}}

\newcommand{\eps}{\varepsilon}
\newcommand{\PF}{\mathfrak{p}}

\newcommand{\g}{\mathfrak{g}}
\newcommand{\red}{\mathrm{Re}D}

\newcommand{\Rm}{{\mathbb R}}
\newcommand{\farc}{\frac}

\begin{document}

\title{The random Schr\"odinger equation: homogenization in
  time-dependent potentials}

\author{Yu Gu\thanks{Department of Mathematics, Building 380, Stanford University, Stanford, CA, 94305, USA (yg@math.stanford.edu; ryzhik@math.stanford.edu)}  \and Lenya Ryzhik\footnotemark[1]}

\maketitle

\begin{abstract}
  We analyze the solutions of the Schr\"odinger equation with the low
  frequency initial data and a time-dependent weakly random
  potential. We prove a homogenization result for the low frequency
  component of the wave field.  We also show that the dynamics
  generates a non-trivial energy in the high frequencies, which do not
  homogenize -- the high frequency component of the wave field remains
  random and the evolution of its energy is described by a kinetic
  equation.  The transition from the homogenization of the low
  frequencies to the random limit of the high frequencies is
  illustrated by understanding the size of the small random
  fluctuations of the low frequency component.
\end{abstract}

\section{Introduction}
 
We consider the Schr\"odinger equation
\begin{equation}
i\partial_t \phi(t,x)+\frac12\Delta \phi(t,x)-\eps V(t,x)\phi(t,x)=0
\label{eq:mainEq}
\end{equation}
with a low frequency initial condition of the form
\begin{equation}\label{june404}
\phi(0,x)=\phi_0(\eps^\alpha x),
\end{equation}
with some $\alpha>0$.  Our goal is to analyze the long time behavior
of $\phi(t,x)$, and understand the energy transfer from the low to
high frequencies that comes about from the inhomogeneities in the
random media.

We define the Fourier transform of $f$ as
\[
\hat f(\xi)=\int_{\Rm^d}e^{-i\xi\cdot x}f(x)dx,
\]
and assume that $V(t,x)$ is a stationary mean-zero Gaussian random
field with a spectral representation
\begin{equation}
V(t,x)=\int_{\R^d}e^{ip\cdot x}\frac{\tilde{V}(t,dp)}{(2\pi)^d}.
\end{equation}
Here $\tilde{V}(t,dp)$ is the stochastic measure and $\tilde{V}(t,dp)=\tilde{V}^*(t,-dp)$, so $V$ is real-valued. Its covariance function and power spectrum are
\[
R(t,x)=\E\{V(s,y)V(s+t,y+x)\},
~~~\hat{R}(\omega,\xi)=\int_{\R^{d+1}}R(t,x)e^{-i\omega t-i\xi\cdot x}dtdx.
\]
The spatial power spectrum (the Fourier transform of $R(t,x)$ in $x$
only)
has the form
\begin{equation}
\tilde{R}(t,\xi)=\int_{\R^d} R(t,x)e^{-i\xi\cdot x}dx=e^{-\mathfrak{g}(\xi)|t|}\hat{R}(\xi),
\end{equation}
where  $\hat{R}(\xi)\in L^1(\R^d)$, and the
spectral gap $\g(\xi)\geq 0$, so that
\begin{equation}
\hat{R}(\omega,\xi)=\frac{2\g(\xi)\hat{R}(\xi)}{\omega^2+\g^2(\xi)}.
\end{equation}
By Bochner's theorem, we have $\hat{R},\tilde{R}\geq 0$.
Throughout the paper, we assume that
\begin{equation}
\frac{\hat{R}(p)}{\g(p)}\in L^1(\R^d) \cap L^\infty(\R^d).
\end{equation}

\subsubsection*{The compensated wave function}

The standard approach to an understanding of the behavior of the
solutions of the weakly random Schr\"odinger equation is in the
context of the kinetic limit~\cite{BPR-NL02,BPR-SD02,BKR-liouv,Spohn,Erdos-Yau2,LS-ARMA-07,BKR-review}, through
the study of the Wigner transform of the solution (the phase space
resolved energy density) \cite{GMMP}.
Our work here is closer in spirit to~\cite{BKR-ARMA-11,KOR-ARMA-13}
that focused not on the weak limit of the energy density of the
solution but on the strong limit of the wave field itself. In order to
motivate the ``correct'' way to this end, let us mention that after a
long time the phase of the wave field acquires a large factor: for
instance, setting $V=0$ in~(\ref{eq:mainEq}) leads to an explicit
expression
\[
\hat\phi(t,\xi)=e^{-i|\xi|^2t/2}\hat\phi(0,\xi)
\]
for the Fourier transform of the solution. Thus, a convenient object in the context of long time behaviors
is the compensated wave function
\begin{equation}\label{june402}
\hat\psi(t,\xi)=e^{i|\xi|^2t/2}\hat\phi(t,\xi),
\end{equation}
which eliminates the deterministic component of the phase.  This
procedure is also known as phase conjugation in the engineering and
physical literature. The surprising miracle is that after this
simple-minded phase compensation, the wave field has a non-trivial
limit.

\subsubsection*{Loose end \#1: the high frequency initial data}

We first describe the results of~\cite{BKR-ARMA-11} obtained when
the initial data for (\ref{eq:mainEq}) is not slowly varying:
\[
\phi(0,x)=\phi_0(x),
\]
that is, $\alpha=0$ in (\ref{june404}). 
Let us set
\begin{equation}
D(p,\xi)=\frac{2\hat{R}(p)}{(2\pi)^d[\g(p)-i(|\xi|^2-|\xi-p|^2)/2]},
~~~~D(\xi)=\int_{\R^d}D(p,\xi)dp.
\end{equation}
It is straightforward to check that 
\begin{equation}
\red(p,\xi)=\frac{2\hat{R}(p)\g(p)}{(2\pi)^d[\g^2(p)+(|\xi|^2-|\xi-p|^2)^2/4]}=\frac{1}{(2\pi)^d}\hat{R}(\frac{|\xi|^2-|\xi-p|^2}{2},p).
\end{equation}
One of the results of~\cite{BKR-ARMA-11} is
that if 
\[
\frac{\hat{R}(p)}{\g(p)}\in L^1(\R^d),
\]
then on the time scale $t\sim\eps^{-2}$,
the compensated wave function corresponding to the initial
data with $\alpha=0$ converges pointwise
in distribution to a Gaussian random variable:
\begin{equation}
\hat{\phi}(\frac{t}{\eps^2},\xi)e^{\frac{i|\xi|^2t}{2\eps^2}}
\Rightarrow \hat{\phi}_0(\xi)e^{-\frac12D(\xi)t}+Z(t,\xi).
\label{eq:bkr}
\end{equation} 
Here, $Z(t,\xi)$ is a centered, complex valued Gaussian with the variance
\begin{equation}
\E\{|Z(t,\xi)|^2\}=\widehat{W}(t,\xi)-|\hat{\phi}_0(\xi)|^2 e^{-\red(\xi) t}.
\end{equation}
The function $\widehat{W}$ solves a (space-homogeneous) kinetic equation
\begin{equation}
\begin{aligned}
\partial_t \widehat{W}
=\int_{\R^d}\hat{R}(\frac{|p|^2-|\xi|^2}{2},p-\xi)(\widehat{W}(t,p)-\widehat{W}(t,\xi))\frac{dp}{(2\pi)^d},
\label{eq:trxi}
\end{aligned}
\end{equation}
with the initial condition 
\[
\widehat{W}(0,\xi)=|\hat{\phi}_0(\xi)|^2.
\]
This result is consistent with the  aforementioned ``traditional" kinetic equation approaches.  
 
\subsubsection*{Loose end \#2: homogenization of the very low
  frequencies}

The results in the high frequency regime ($\alpha=0$) should be
contrasted with the analysis of Bal and Zhang
in~\cite{ZB-CMS-14,ZB-SD-14} for the case $\alpha=1$ in
(\ref{june404}), performed for time-independent potentials. For the initial value problem
\begin{eqnarray}\label{june406}
&&i\phi_t+\frac{1}{2}\Delta\phi-\eps V(x)\phi=0,\\
&&\phi(0,x)=\phi_0(\eps x),\nonumber
\end{eqnarray}
with a mean-zero Gaussian random potential $V(x)$,
they have established a homogenization result:
\[
\phi^\eps(t,x):=\phi\Big(\frac{t}{\eps^2},\frac{x}{\eps}\Big)
\]
converges in probability, as $\eps\to 0$ to a deterministic limit
$\bar\phi(t,x)$, which satisfies the Schr\"odinger equation
\begin{eqnarray}\label{june408}
&&i\bar\phi_t+\frac{1}{2}\Delta\bar\phi-\bar V\bar\phi=0,\\
&&\bar\phi(0,x)=\phi_0(x).\nonumber
\end{eqnarray}
The effective potential is constant and is given by
\[
\bar V=\int_{\R^d}\frac{\hat R(p)dp}{|p|^2}.
\]
Let us mention that the choice $\alpha=1$ is special, as then the overall
phase of the solution at the times $t\sim\eps^{-2}$ is 
\[
\frac{t}{\eps^2}\eps^2|\xi|^2=O(1),
\]
so that no phase compensation is needed.  

\subsection*{Homogenization of the low frequencies}

Summarizing the above results, while solutions of (\ref{eq:mainEq})
with the high frequency initial data have a random limit on the time
scale $t\sim\eps^{-2}$, as in (\ref{eq:bkr}), solutions with the
``very slowly varying'' initial data as in (\ref{june406}) are
homogenized on this time scale -- their limit is deterministic.  The
first goal of this paper is to understand where the transition between
the two regimes occurs -- this is the motivation for introducing a
general $\alpha>0$ in (\ref{june404}). It will turn out that the
homogenization result (formulated for the compensated wave function)
holds for all $\alpha>0$ -- that is, no matter how ``relatively high"
the low frequency of the initial condition is, solution has a
deterministic limit at times $t\sim\eps^{-2}$.  However, we will see
that, unlike in the setting of~\cite{ZB-CMS-14,ZB-SD-14}, the temporal
fluctuations of the random potential lead to an effective potential
with a non-trivial imaginary part. This means that the homogenized
field loses mass in the limit. This loss of mass is attributed to the
energy transfer to the high frequencies, which, as we show, account
for the mass missing in the low frequencies, do not homogenize, and
satisfy a kinetic type limit.  We also analyze the random fluctuations
of the low frequency component of the wave field and characterize the
corrector to the homogenized limit.

More precisely, we consider the Schr\"odinger equation
\begin{equation}
i\partial_t \phi(t,x)+\frac12\Delta \phi(t,x)-\eps V(t,x)\phi(t,x)=0
\label{eq:mainEqbis}
\end{equation}
with a low frequency initial condition 
\begin{equation}\label{june502}
\phi(0,x)=\phi_0(\kappa x),
\end{equation} 
with $\kappa\ll 1$. 
The Fourier transform of the initial condition is
\[
\hat\phi(0,\xi)=\kappa^{-d}\hat{\phi}_0\big(\farc{\xi}{\kappa}\big).
\]
Thus, if the function $\hat\phi_0(\xi)$ is of the Schwartz class, 
$\hat\phi(0,\xi)$ is concentrated on the wave vectors $\xi$ of the
size $O(\kappa)$. While the Schr\"odinger equation
with a time-dependent potential conserves the total mass:
\begin{equation}\label{apr2714}
M(t)=\int_{\Rm^d}|\phi(t,x)|^2dx=\int_{\Rm^d}|\phi(0,x)|^2dx,
\end{equation}
the total energy 
\begin{equation}\label{apr2716}
E(t)=\int_{\Rm^d}[|\nabla\phi|^2+\eps V|\phi|^2]dx
\end{equation}
is not conserved, unlike for time-independent potentials.  Thus, even
if the mass is initially concentrated in the low wave numbers,  
after a long time evolution it may spread
to $O(1)$ frequencies as well. As the potential is weak, the time it
takes for the mass to spread over a range of frequencies will be long.

%
We consider the long time behavior of the solution, on the time scale
of the order $t\sim\eps^{-2}$, when the effect of the weak random
potential will be non-trivial.  We will first consider the ``low
frequency'' rescaled compensated wave function:
\begin{equation}\label{apr2722}
\psi_\eps(t,\xi)=
\kappa^{d}\hat{\phi}(\frac{t}{\eps^2},\kappa\xi)e^{\frac{i\kappa^2|\xi|^2t}{2\eps^{2}}}
\end{equation}
with the initial data $\psi_\eps(0,\xi)=\hat{\phi}_0(\xi)$.  This allows us to study the low frequency
component of the solution -- wave numbers of the order $O(\kappa)$. A
straightforward computation shows that this function is a solution of
the following integral equation
\begin{equation}
\psi_\eps(t,\xi)=\hat{\phi}_0(\xi)+\frac{1}{i\eps}\int_0^t \int_{\R^d}
\frac{\tilde{V}(\frac{s}{\eps^2},dp)}{(2\pi)^d}
e^{i\kappa^2(|\xi|^2-|\xi-\frac{p}{\kappa}|^2)\frac{s}{2\eps^{2}}}
\psi_\eps(s,\xi-\frac{p}{\kappa})ds.
\label{eq:mainEq1}
\end{equation}
We have the following result  for the low frequencies.
\begin{theorem}\label{thm:homo}
Assume that $\kappa=\eps^\alpha$ with $\alpha>0$. Then, 
for fixed $t>0$ and $\xi\in\R^d$, 
\begin{equation}\label{apr2708}
\hbox{$\psi_\eps(t,\xi)\to \bar\psi(t,\xi)=\hat{\phi}_0(\xi)e^{-\frac12D(0)t}$ 
in probability as $\eps\to 0$.}
\end{equation}
\end{theorem}
Let us stress that $\xi=O(1)$ in the argument of the
function $\psi_\eps(t,\xi)$ corresponds to $\xi=O(\kappa)$ in the
argument of the function $\phi$ -- Theorem~\ref{thm:homo} 
addresses the evolution of the low frequencies of the solution of the Schr\"odinger equation with
a slowly varying initial condition.  
Recall that 
\begin{equation}
D(0)=\int_{\R^d}\frac{2\hat{R}(p)}{(2\pi)^d(\g(p)+i|p|^2/2)}dp,
\label{eq:homocon}
\end{equation}
and, as $\g(p)\ge 0$, we have $\red(0)>0$. Therefore, the passage to
limit $\eps\to 0$ in (\ref{apr2708}) induces a loss of the $L^2(\R^d)$
norm: while
\[
\|\psi_\eps(t,\cdot)\|_{L^2}=\|\phi_0\|_{L^2},
\]
as can be seen simply from the definition of $\psi_\eps(t,\xi)$, we
have 
\[
\|\bar\psi(t,\cdot)\|_{L^2}=\|{\phi}_0\|_{L^2}e^{-\red(0)t/2}<\|\phi_0\|_{L^2}.
\]
The natural question is how does the loss of mass happen, and where
does the mass go?  Mathematically, there is no contradiction, as we
will show the convergence in Theorem~\ref{thm:homo} is not uniform
with respect to $\xi\in\R^d$. From a physical point of view, as we
have mentioned, the time dependence of the random potential
breaks the conservation of the energy (\ref{apr2716}), which allows
the mass to escape to the high frequencies. Let us mention that in the
\emph{time-independent} case \cite{bckr}, where the conservation of
the energy prevents the escape of mass from the low frequencies, it is
shown that the mass is conserved as well.

\subsection*{Generation of the high frequencies}

We now investigate the generation of the high frequencies in the above
setting. Once again, we consider the solution $\phi(t,x)$ of
(\ref{eq:mainEqbis}) with the initial data (\ref{june502}). We stress
that in all our results the initial condition (\ref{june502}) is the
same -- various rescalings in Theorem~\ref{thm:homo} above and
Theorems~\ref{thm:trHigh},~\ref{thm:cor} and~\ref{thm:wig} below
correspond to zooming into various frequency ranges in the same
solution.  Our next goal is to understand how the mass escapes from the low
frequencies (those of the initial condition) to the high frequencies,
generated by the interaction with the random potential.
As we are now interested in the high and not the low frequencies, we
define the compensated wave function not quite as in (\ref{apr2722}),
but as
\begin{equation}\label{apr2724}
\Psi_\eps(t,\xi)=\kappa^{\frac{d}{2}}\hat{\phi}(\frac{t}{\eps^2},\xi)e^{\frac{i|\xi|^2t}{2\eps^{2}}},
\end{equation}
so that the frequency is not rescaled.
The initial condition for $\Psi_\eps$ is
\[
\Psi_\eps(0,\xi)=\kappa^{-d/2}\hat{\phi}_0(\xi/\kappa).
\]
The pre-factor $\kappa^{d/2}$ in (\ref{apr2724}) is chosen so that we get a non-trivial limit. This function solves the integral equation
\begin{equation}
\Psi_\eps(t,\xi)=\frac{1}{\kappa^{d/2}}\hat{\phi}_0(\frac{\xi}{\kappa})+
\frac{1}{i\eps}\int_0^t \int_{\R^d} \frac{\tilde{V}(\frac{s}{\eps^2},dp)}{(2\pi)^d}e^{i(|\xi|^2-|\xi-p|^2)\frac{s}{2\eps^{2}}}\Psi_\eps(s,\xi-p)ds.
\label{eq:mainEq2}
\end{equation}
The following result explains the loss of mass observed in Theorem~\ref{thm:homo}, and tracks the generation of the high frequencies.
\begin{theorem}\label{thm:trHigh}
Assume that $\kappa=\eps^\alpha$ with $\alpha>0$, then for fixed $t>0$ and $\xi\neq 0$, 
we have 
\[
\hbox{$\Psi_\eps(t,\xi)\Rightarrow \bar Z(t,\xi)$ in law as $\eps\to 0$,}
\]
where $\bar Z(t,\xi)$ is a centered, complex valued Gaussian random variable. Its 
variance $\widehat{W}_\delta(t,\xi)$ is the solution of \eqref{eq:trxi} 
with the initial condition $\widehat{W}_\delta(0,\xi)=\|\hat{\phi}_0\|_2^2\delta(\xi)$.
 \end{theorem}

The variance $\widehat{W}_\delta(t,\xi)$ can be explicitly written as a series expansion 
\begin{equation}
\widehat{W}_{\delta}(t,\xi)=\widehat{W}_{\delta,b}(t,\xi)+\widehat{W}_{\delta,s}(t,\xi),
\end{equation}
 with the ballistic part 
\begin{equation*}
\widehat{W}_{\delta,b}(t,\xi)=\|\hat{\phi}_0\|^2e^{-\red(0)t}\delta(\xi),
\end{equation*}
and the scattering part
\begin{equation*}
\begin{aligned}
\widehat{W}_{\delta,s}(t,\xi)=\sum_{k=1}^\infty
\|\hat{\phi}_0\|_2^2\int_{0=v_{k+1}\leq v_k\leq \ldots\leq v_1\leq v_0=t} dv&\int_{\R^{kd}} dP
\prod_{j=0}^{k}e^{-(v_j-v_{j+1})\red(\xi-\ldots-P_j)}\\
\times &
\prod_{j=1}^{k} \red(P_j,\xi-\ldots-P_{j-1})\delta(\xi-P_1-\ldots-P_{k}).
\end{aligned}
\end{equation*}
Let us mention that $\widehat{W}_\delta(t,\xi)=\widehat{W}_{\delta,s}(t,\xi)$ when $\xi\neq 0$, that is,
only the scattering part contributes to the variance in Theorem~\ref{thm:trHigh}. 
We also observe 
\[
\int_{\R^d}\widehat{W}_{\delta,b}(t,\xi)d\xi=\|\hat{\phi}_0\|_2^2e^{-\red(0)t},
\]
which equals to the mass of the low frequency waves.

Theorems~\ref{thm:homo} and \ref{thm:trHigh} describe the dynamics of
\eqref{eq:mainEq} on different scales of the frequency domain. In the
former case, the low frequencies are zoomed in, and we find a
deterministic evolution (homogenzation).  In the latter, we track the
high frequency component of the solution, so that the low frequency
initial condition shrinks to a point source at the origin, which
generates the high frequency waves.

\subsection*{The fluctuation analysis in homogenization regime}

We now return to the analysis of the behavior of the low
frequencies. According to Theorem~\ref{thm:homo}, the compensated wave
function homogenizes for the low frequencies, hence the next interesting
object is the fluctuation, which we define as
\[
\U_\eps(t,\xi)=\frac{1}{\kappa^{d/2}}(\psi_\eps(t,\xi)-\E\{\psi_\eps(t,\xi)\}).
\]
Here, $\psi_{\eps}(t,\xi)$ is defined as in (\ref{apr2722}).
Heuristically, since the homogenization limit in
Theorem~\ref{thm:homo} captures the ballistic component of the wave
field, we expect small random fluctuations consisting of the remaining
scattering components.  Indeed, we will see that the fluctuation
exhibits a kinetic-like behavior.  Let us set
\begin{equation}
\mathcal{W}_\alpha(t,\xi)=\left\{
\begin{array}{ll}
0 & \mbox {if }\alpha\in (0,1),\\
-D(0,0)e^{-D(0)t}\displaystyle\int_0^t \int_{\R^d}\hat{\phi}_0(\xi-p)\hat{\phi}_0(\xi+p)e^{-i|p|^2v}dpdv &  \mbox {if } \alpha=1,\\
-D(0,0)te^{-D(0)t} \displaystyle\int_{\R^d}\hat{\phi}_0(\xi-p)\hat{\phi}_0(\xi+p)dp &  \mbox {if } \alpha>1.
 \end{array}
 \right.
\end{equation}

\begin{theorem} Assume that $\kappa=\eps^\alpha$, then
for fixed $t>0$ and $\xi\in\R^d$, we have
\[
\hbox{$\U_\eps(t,\xi)\Rightarrow Z_\delta(t,\xi)=X_\delta(t,\xi)+iY_\delta(t,\xi)$ as $\eps\to 0$,}
\]
 where $X_\delta,Y_\delta$ are centered, jointly Gaussian random variables such that 
 \[
 \E\{|Z_\delta(t,\xi)|^2\}=\widehat{W}_{\delta,s}(t,0),
 \]
 and 
 \[
 \E\{Z_\delta(t,\xi)^2\}=\mathcal{W}_\alpha(t,\xi).
 \]
\label{thm:cor}
\end{theorem}
Therefore, we can write 
\[
\psi_\eps(t,\xi)=\E\{\psi_\eps(t,\xi)\}+\kappa^{d/2}\U_\eps(t,\xi),
\]
and Theorem~\ref{thm:cor} shows that when $\kappa=\eps^{\alpha}$, with
$\alpha<1$, the fluctuation $\U_\eps(t,\xi)$ is approximately
distributed as $Z_{\delta}(t,0)$, a centered complex Gaussian random
variable with variance $\widehat{W}_{\delta,s}(t,0)$. This is similar
to the result of Theorem~\ref{thm:trHigh} for the high frequency,
albeit the variance is now given by the transport solution evaluated
at the origin $\xi=0$, since we are now in the low frequency regime.
If we let $\alpha\to 0$ (which is the same as $\kappa\to 1$, so that
the initial condition is less and less slowly varying), then,
formally, $\psi_\eps(t,\xi)$ is distributed as
\[
\hat{\phi}_0(\xi)e^{-\frac12D(0)t}+Z_\delta(t,0),
\]
which is consistent with \eqref{eq:bkr}. That is,
Theorem~\ref{thm:cor} also interpolates between the deterministic
limit for the low frequencies and the random behavior of the high
frequency component of the solution. 

\subsubsection*{The Wigner transform of the random fluctuation}

Besides the pointwise fluctuation for a fixed $\xi\in \R^d$, we also
consider the fluctuation of $\psi_\eps(t,\xi)$ as a wave field. The
tool we use is the Wigner transform for some $\beta\geq 0$:
\begin{equation}
W_\eps(t,x,\xi)=\int_{\R^d} \U_\eps(t,\xi+\frac{\eps^\beta\eta}{2})\U_\eps^*(t,\xi-\frac{\eps^\beta\eta}{2})e^{i\eta\cdot x}\frac{d\eta}{(2\pi)^d}.
\end{equation}
Let $\bar{W}_{\delta}$ be the solution to  the kinetic equation
\begin{equation}
 \partial_t \bar{W}+\xi\cdot \nabla_x \bar{W}=\int_{\R^d}\hat{R}(\frac{|p|^2-|\xi|^2}{2},p-\xi)(\bar{W}(t,x,p)-\bar{W}(t,x,\xi))\frac{dp}{(2\pi)^d},
\label{eq:trxxi}
\end{equation}
%
with
the initial condition
\[
\bar{W}_\delta(0,x,\xi)=\|\hat{\phi}_0\|_2^2\delta(\xi)\delta(x),
\]
and $\bar{W}_{\delta,b},\bar{W}_{\delta,s}$ be the ballistic and
scattering component of $\bar W_\delta$, respectively:
\begin{equation*}
\bar{W}_{\delta,b}(t,x,\xi)=\|\hat{\phi}_0\|_2^2\delta(\xi)\delta(x)e^{-\red(0)t},
\end{equation*}
and
\begin{equation*}
\begin{aligned}
\bar{W}_{\delta,s}(t,x,\xi)=\sum_{k=1}^\infty
\|\hat{\phi}_0\|_2^2&\int_{0=v_{k+1}\leq v_k\leq \ldots\leq v_1\leq v_0=t} dv\int_{\R^{kd}} dP
\prod_{j=0}^{k}e^{-(v_j-v_{j+1})\red(\xi-\ldots-P_j)}\\
\times &
\prod_{j=1}^{k} \red(P_j,\xi-\ldots-P_{j-1})\delta(\xi-P_1-\ldots-P_{k})\delta(x-\xi t+\sum_{j=1}^k P_jv_j).
\end{aligned}
\end{equation*}

\begin{theorem}
Assume that $\kappa=\eps^{\alpha}$, $\alpha\in (0,1)$ and
$\alpha+\beta=2$, 
then for any test function $\varphi\in \mathcal{S}(\R^{2d})$ and $t>0$, 
\[
\int_{\R^{2d}}W_\eps(t,x,\xi)\varphi^*(x,\xi)dxd\xi \to \int_{\R^{2d}} \bar{W}_{\delta,s}(t,x,0)\varphi^*(x,\xi)dxd\xi
\] in probability as $\eps\to 0$.
\label{thm:wig}
\end{theorem}
As Theorem~\ref{thm:homo} indicates that the ballistic component of
transport solution gives the low frequency behavior, we conclude from
Theorems~\ref{thm:cor} and \ref{thm:wig} that the small random
fluctuations are described by the scattering component of the solution
of the kinetic equation.  

This paper is organized as follows. First, in
Section~\ref{sec:duhamel} we present the Duhamel expansion and the
corresponding diagrammatic expansions and the moment estimates that
are needed for the proofs of all theorems.
Section~\ref{sec:homlow} contains the proof of
Theorem~\ref{thm:homo}. Theorem~\ref{thm:trHigh} is proved in
Section~\ref{sec:high}. Finally, Theorems~\ref{thm:cor}
and~\ref{thm:wig} are proved in Section~\ref{sec:fluct}.

{\bf Acknowledgment.} This work was supported by an AFOSR NSSEFF Fellowship and
NSF grant DMS-1311903.

\section{The Duhamel expansion and the moment estimates}\label{sec:duhamel}

Theorems \ref{thm:homo}, \ref{thm:trHigh}, \ref{thm:cor} and
\ref{thm:wig} are all proved using the moment method. For the
convergence 
\[
\psi_\eps(t,\xi)\to \hat{\phi}_0(\xi)e^{-\frac12D(0)t},
\]
in probability (Theorem~\ref{thm:homo}), it suffices to show the
convergence of $\E\{\psi_\eps(t,\xi)\}$ and
$\E\{|\psi_\eps(t,\xi)|^2\}$ to their respective limits. For the
convergence in law of $\Psi_\eps(t,\xi)$ and $\U_\eps(t,\xi)$ to a
Gaussian in Theorems~\ref{thm:trHigh} and~\ref{thm:cor}, respectively,
we need to show the convergence of the corresponding
moments~$\E\{\Psi_\eps(t,\xi)^M(\Psi_\eps^*(t,\xi))^N\}$
and~$\E\{\U_\eps(t,\xi)^M(\U_\eps^*(t,\xi))^N\}$ for any $M,N\in
\mathbb{N}$ to their respective limits, which makes the analysis
slightly more computationally heavy. In this section, we perform the
preliminary moment estimates that are needed in the proofs of the
theorems.

\subsubsection*{The Duhamel expansions}

All moment estimates rely on the Duhamel expansions that we now
recall. From now on, we will set~$\kappa=\eps^\alpha$.
For the low frequencies, we can iterate the integral equation
\eqref{eq:mainEq1} for the function~$\psi_\eps(t,\xi)$, and write the
solution as a series
\begin{equation}
\psi_\eps(t,\xi)=\sum_{n=0}^\infty f_{n,\eps}(t,\xi),
\label{eq:lowSo}
\end{equation}
with the individual terms
\begin{equation}
f_{n,\eps}(t,\xi)=\frac{1}{(i\eps)^n}\int_{\Delta_n(t)}
\int_{\R^{nd}}\prod_{j=1}^n 
\frac{\tilde{V}(\frac{s_j}{\eps^2},dp_j)}{(2\pi)^d}
e^{iG_n(\eps^\alpha \xi, s^{(n)},p^{(n)})/\eps^{2}} 
\hat{\phi}_0(\xi-\frac{p_1+\ldots+p_n}{\eps^\alpha}),
\label{eq:homoTerm}
\end{equation}
and the phase factor 
\begin{equation}
G_n(\xi, s^{(n)},p^{(n)})=
\sum_{k=1}^n (|\xi-p_1-\ldots-p_{k-1}|^2-|\xi-p_1-\ldots-p_k|^2)\frac{s_k}{2}.
\end{equation}
Here, we used the convention $f_{0,\eps}(t,\xi)=\hat{\phi}_0(\xi)$,
and have set $p_0=0$, $p^{(n)}=(p_1,\ldots,p_n)$,
as well as~$s^{(n)}=(s_1,\ldots,s_n)$.
We have also defined the time simplex
\[
\Delta_n(t)=\{0\leq s_n\leq\ldots\leq s_1\leq t\}.
\]

For the high frequencies, the solution $\Psi_\eps(t,\xi)$ 
to \eqref{eq:mainEq2} is
similarly written as
\begin{equation}
\Psi_\eps(t,\xi)=\sum_{n=0}^\infty F_{n,\eps}(t,\xi)
\label{eq:highSo}
\end{equation} with
\begin{equation}
F_{n,\eps}(t,\xi)=\frac{1}{(i\eps)^n}
\int_{\Delta_n(t)}\int_{\R^{nd}}\prod_{j=1}^n 
\frac{\tilde{V}(\frac{s_j}{\eps^2},dp_j)}{(2\pi)^d}
e^{iG_n( \xi, s^{(n)},p^{(n)})/\eps^{2}} 
\frac{1}{\eps^{\alpha d/2}}\hat{\phi}_0(\frac{\xi-p_1-\ldots-p_n}{\eps^\alpha})
\label{eq:kineTerm}
\end{equation}
and
\[
F_{0,\eps}=\frac{1}{\eps^{\alpha
    d/2}}\hat{\phi}_0(\frac{\xi}{\eps^\alpha}).
\] 
The key ``bureaucratic'' difference between the Duhamel expansions
\eqref{eq:homoTerm} and \eqref{eq:kineTerm} for the functions~$\psi_\eps(t,\xi)$ and
$\Psi_\eps(t,\xi)$ is that $\eps^\alpha\xi\mapsto \xi$. This will
make the limits very different.

The following lemma ensures that the solutions given by
\eqref{eq:lowSo} and \eqref{eq:highSo} are well-defined and we can
interchange the summation and the expectation when computing the
moments. Its proof is exactly as that of \cite[Proposition 3.8]{BKR-ARMA-11}.
\begin{lemma}
Fix $\eps>0,M,N\in\mathbb{N}$. Let $g_{n,\eps}=f_{n,\eps}$ or $F_{n,\eps}$, then
\begin{equation}
|\E\{g_{m_1,\eps}\ldots g_{m_M,\eps} g_{n_1,\eps}^*\ldots
g_{n_N,\eps}^*\}|
\leq C_\eps(m_1,\ldots,m_M,n_1,\ldots,n_N)
\end{equation}
with 
\[
\sum_{m_1,\ldots,m_M=0}^\infty \sum_{n_1,\ldots,n_N=0}^\infty
C_\eps(m_1,\ldots,m_M,n_1,\ldots,n_N)<\infty.
\]
\label{lem:gnbd}
\end{lemma}


\subsubsection*{The pairings}


Now, we discuss in detail the calculation of the moments
\[
\E\{g_{m_1,\eps}\ldots g_{m_M,\eps} g_{n_1,\eps}^*\ldots
g_{n_N,\eps}^*\},
\]
where $g_{n,\eps}=f_{n,\eps}$ or $F_{n,\eps}$, and
\[
\sum_{i=1}^M m_i+\sum_{j=1}^N n_j=2k,
\]
for some $k\in\mathbb{N}$ (if
the sum is odd, then the moment is zero by the Gaussian
property). 
We have
\begin{equation}
\begin{aligned}
&\E\{g_{m_1,\eps}\ldots g_{m_M,\eps} g_{n_1,\eps}^*\ldots
g_{n_N,\eps}^*\} 
={(i\eps)^{-\sum_{i=1}^Mm_i}}{(-i\eps)^{-\sum_{j=1}^Nn_j}}\\
&~~~~~~~~~~~~~~~~~~~\times\int_{\Delta_{m_1}(t)\times\ldots\times\Delta _{n_N}(t)} dsdu
\int_{\R^{2kd}} 
\E\{I_{M,N}\} e^{i\G_M}e^{-i\G_N}\prod_{i=1}^M  h_{M,i}\prod_{j=1}^N h_{N,j}^*,
\end{aligned}
\label{eq:mm}
\end{equation}
with 
\begin{equation*}
\begin{aligned}
I_{M,N}=&\farc{1}{(2\pi)^{2kd}}\tilde{V}(\frac{s_{1,1}}{\eps^2},dp_{1,1})\ldots 
\tilde{V}(\frac{s_{1,m_1}}{\eps^2},dp_{1,m_1})\ldots 
\tilde{V}(\frac{s_{M,1}}{\eps^2},dp_{M,1})\ldots
\tilde{V}(\frac{s_{M,m_M}}{\eps^2},dp_{M,m_M})\\
&\times\tilde{V}^*(\frac{u_{1,1}}{\eps^2},dq_{1,1})\ldots 
\tilde{V}^*(\frac{u_{1,n_1}}{\eps^2},dq_{1,n_1})\ldots 
\tilde{V}^*(\frac{u_{N,1}}{\eps^2},dq_{N,1})
\ldots\tilde{V}^*(\frac{u_{N,n_N}}{\eps^2},dq_{N,n_N}),
\end{aligned}
\end{equation*}
and the phases
\[
\G_M=\sum_{i=1}^MG_{m_i}(\eta, s^{(m_i)}_i,p^{(m_i)}_i)/\eps^2,
~~~\G_N=\sum_{i=1}^NG_{n_i}(\eta, u^{(n_i)}_i,q^{(n_i)}_i)/\eps^2,
\]
with $\eta=\eps^\alpha\xi$ or $\xi$, depending on whether
$g_{n,\eps}=f_{n,\eps}$ or $F_{n,\eps}$.  The initial conditions
appear as
\[
h_{M,i}=\hat{\phi}_0(\xi-\frac{p_{i,1}+\ldots+p_{i,m_i}}{\eps^\alpha}),
~~~
h_{N,j}^*=\hat{\phi}_0^*(\xi-\frac{q_{j,1}+\ldots+q_{j,n_j}}{\eps^\alpha})
\]
when $g_{n,\eps}=f_{n,\eps}$, and as 
\[
h_{M,i}=\eps^{-\alpha d/2}
\hat{\phi}_0(\frac{\xi-p_{i,1}-\ldots-p_{i,m_i}}{\eps^\alpha}),
~~~
h_{N,j}^*=\eps^{-\alpha d/2}
\hat{\phi}_0^*(\frac{\xi-q_{j,1}-\ldots-q_{j,n_j}}{\eps^\alpha}),
\]
when $g_{n,\eps}=F_{n,\eps}$.

Using the rules of computing the $2k-$th joint moment of 
mean zero Gaussian random variables, we obtain
\begin{equation}
\E\{I_{M,N}\}=\sum_{\F} \prod_{(v_l,v_r)\in\F}(2\pi)^{-d} e^{-\g(w_l)|v_{l}-v_{r}|/\eps^2}\delta(w_l+ w_r)\hat{R}(w_l)dw_ldw_r.
\label{eq:IMN}
\end{equation}
The summation $\sum_\F$ is taken over all allocations of the set of the vertices
\[
\{s_{1,1},\ldots,s_{1,m_1},\ldots,s_{M,1},\ldots,s_{M,m_M}u_{1,1},
\ldots,u_{1,n_1},\ldots,u_{N,1},\ldots,u_{N,n_N}\}.
\] 
into $k$ (unordered) pairs (recall that $\sum_{i=1}^M m_i+\sum_{j=1}^N n_j=2k$). We call each allocation a pairing. In (\ref{eq:IMN}), $v_l,v_r$ are the two vertices of a given pair, and
$w_l,w_r$ are the respective $p,q$ variables, that is, $w_l=p_{i,j}$ if
$v_l=s_{i,j}$ and $w_l=-q_{i,j}$ if $v_l=u_{i,j}$. The same holds for
$w_r$. We will also write a pair as an edge $e=(v_l,v_r)$. Note that the
order of $v_l,v_r$ does not matter here since $\g,\hat{R}$ are both
even. 

\subsubsection*{A uniform bound on the pairings}

We recall the following general bound.
\begin{lemma}
Let $g_{n,\eps}=f_{n,\eps}$ or $F_{n,\eps}$, then we have, for all $\eps\in(0,1]$,
\begin{equation}
|\E\{g_{m_1,\eps}\ldots g_{m_M,\eps} g_{n_1,\eps}^*\ldots g_{n_N,\eps}^*\}|\leq\frac{(2k-1)!!}{\prod_{i=1}^M (m_i)!\prod_{j=1}^N (n_j)!}C^k
\end{equation}
with some constant $C$ depending on $t,\xi,\hat{R},\g$. Here $2k=\sum_{i=1}^M m_i+\sum_{j=1}^N n_j$.
\label{lem:bdmm}
\end{lemma}
{\bf Proof.}
The proof is close to the case $g=f_{n,\eps}$ and $\alpha=0$ which is
already contained in~\cite{BKR-ARMA-11}.  We present it, together with the
required modifications, for the convenience of the reader.
%
By symmetry, the RHS of \eqref{eq:mm} can be bounded by
\begin{equation}
 \frac{1}{\prod_{i=1}^M (m_i)!\prod_{j=1}^N (n_j)!}
\frac{1}{\eps^{2k}}\int_{[0,t]^{2k}} dsdu \int_{\R^{2kd}} |\E\{I_{M,N}\}| \prod_{i=1}^M  |h_{M,i}|\prod_{j=1}^N |h_{N,j}^*|.
 \label{eq:mmbd2}
\end{equation}
In the case when $g_{n,\eps}=f_{n,\eps}$, we bound 
\[
\prod_{i=1}^M  |h_{M,i}|\prod_{j=1}^N |h_{N,j}^*|\leq
\|\hat{\phi}_0\|_\infty^{M+N},
\]
then for a given pairing $\F$, we have
\[
\frac{1}{\eps^{2k}}\int_{[0,t]^{2k}} dsdu \int_{\R^{2kd}}\prod_{(v_l,v_r)\in\F}(2\pi)^{-d} 
e^{-\g(w_l)|v_{l}-v_{r}|/\eps^2}\delta(w_l+ w_r)\hat{R}(w_l)dw_ldw_r\leq C^k,
\]
where we used the integrability of $\hat{R}(p)/\g(p)$. Thus,
\eqref{eq:mmbd2} can be bounded by
\begin{equation}
\frac{\#(\F)}{\prod_{i=1}^M (m_i)!\prod_{j=1}^N (n_j)!}C^k\|\hat{\phi}_0\|_\infty^{M+N}=
\frac{(2k-1)!!}{\prod_{i=1}^M (m_i)!\prod_{j=1}^N (n_j)!}C^k\|\hat{\phi}_0\|_\infty^{M+N}.
\end{equation}

In the case when $g_{n,\eps}=F_{n,\eps}$, we consider $\xi\neq 0$ and integrate $w_r$ and bound
\eqref{eq:mmbd2} by
\begin{equation}
 \frac{1}{\prod_{i=1}^M (m_i)!\prod_{j=1}^N (n_j)!}\frac{1}{\eps^{2k}}\int\limits_{[0,t]^{2k}} \!\!dsdu 
 \int\limits_{\R^{kd}} \sum_{\F} \prod_{(v_l,v_r)\in\F}  \!\!\!
 e^{-\g(w_l)|v_{l}-v_{r}|/\eps^2}\hat{R}(w_l) \prod_{i=1}^M  |h_{M,i}|\prod_{j=1}^N |h_{N,j}^*|
 \prod_{l=1}^k\frac{dw_l}{(2\pi)^d}.
 \label{eq:mmbd3}
 \end{equation}
For a given pairing $\F$, we have
\begin{eqnarray}\label{apr3006}
|h_{M,i}|=\eps^{-\alpha
   d/2}|\hat{\phi}_0|\big(\frac{P_i}{\eps^\alpha}\big), ~~~
|h_{N,j}^*|=\eps^{-\alpha d/2}|\hat{\phi}_0|\big(\frac{Q_j}{\eps^\alpha}\big),
\end{eqnarray}
where 
\[
P_i=\xi-p_{i,1}-\ldots-p_{i,m_i},
~~~Q_j=\xi-q_{j,1}-\ldots-q_{j,n_j},
\]
subject to the conditions 
\begin{equation}\label{apr3002}
\hbox{$w_l+w_r=0$ when $(v_l,v_r)\in \F$. }
\end{equation}
The difference with the previous case are the factors
$\eps^{-\alpha d/2}$ in (\ref{apr3006}). Note that if $P_i=\xi$ or~$Q_j=\xi$ (this may
happen because of (\ref{apr3002})), as $\xi\neq 0$ is fixed and
$\hat{\phi}_0$ is rapidly decaying, we may simply use the bound
\[
\eps^{-\alpha d/2}|\hat{\phi}_0|\big(\frac{\xi}{\eps^\alpha}\big)\leq C.
\]
For $i$, $j$ such that
$P_i,Q_j\neq\xi$, to deal with the large factors in
(\ref{apr3006}),
we change variables as follows. Take some $i$ with $P_i\neq \xi$, so that
\[
p_{i,1}+\ldots+p_{i,m_i}\neq 0.
\]
We pick any variable $p$ from $\{p_{i,1},\ldots,p_{i,m_i}\}$ (note the
number of elements here can be strictly smaller than $m_i$ since we
have already integrated out the variables $w_r$), and change $p$ to
$p'=P_i/\eps^\alpha$. The variable $p=w_l$ was paired to some
$p_j$ or $q_j=w_r$ as in (\ref{apr3002}).  Thus, after the integration of
$w_r$, $p'$ will also appear in a unique $\tilde{h}_{M,i}$ which
equals to some $h_{M,j}$ or $h_{N,j}^*$.  We use the bound
\[
|\tilde{h}_{M,i}|\leq \eps^{-\alpha d/2}C.
\] 
Thus, after the change of variable and taking into account the
Jacobian of the change of variables, we have, with a slight abuse of
notation
\begin{equation}\label{apr3010}
|h_{M,i}\tilde{h}_{M,i}|dp\leq \eps^{-\alpha
  d/2}|\hat{\phi}_0|(p')\eps^{-\alpha d/2}C \eps^{\alpha d}dp'=
C|\hat{\phi}_0|(p')dp'.
\end{equation}
Since the change of variable only relates to $p_i$, all other
$h_{M,i},h_{N,j}^*$ are not affected. We continue the
procedure, integrating out the $p$-variables one by one. 
If we are left with a single
\[
\hbox{$|h|=\eps^{-\alpha d/2}|\hat{\phi}_0|(P_i/\eps^\alpha)$ or
$\eps^{-\alpha d/2}|\hat{\phi}_0|(Q_i/\eps^\alpha)$}
\]
in the end,  we change variable similarly, and estimate this term,
together with the Jacobian as
\begin{equation}\label{apr3012}
\eps^{\alpha d/2}|\hat{\phi}_0|(p')dp'\leq |\hat{\phi}_0|(p')dp'.
\end{equation}
Overall, this change of variables will involve $M+N$ momenta, and will
eliminate all factors $h_{M,i}$ and $h_{N,j}^*$, and we
will be left with an expression of the form
\begin{equation}
\begin{aligned}
&\frac{1}{\eps^{2k}}\int_{[0,t]^{2k}} dsdu \int_{\R^{kd}}
\prod_{(v_l,v_r)\in\F}(2\pi)^{-d} 
e^{-\g(w_l)|v_{l}-v_{r}|/\eps^2}\hat{R}(w_l) \prod_{i=1}^M  |h_{M,i}|\prod_{j=1}^N |h_{N,j}^*|dw\\
\leq & \frac{C^{M+N}}{\eps^{2k}}\int_{[0,t]^{2k}} dsdu 
\int_{\R^{kd}} \prod_{(v_l,v_r)\in\F_1}(2\pi)^{-d}
e^{-\g(w_l)|v_{l}-v_{r}|/\eps^2}\hat{R}(w_l)
\prod_{(v_l,v_r)\in\F_2}(2\pi)^{-d} e^{-\g(z_l)|v_{l}-v_{r}|/\eps^2}\hat{R}(z_l)\\
 &\times\prod_{(v_l,v_r)\in\F_2}|\hat{\phi}_0|(w_l)dw.
 \label{eq:mbcv}
\end{aligned}
\end{equation}
Here, $(v_l,v_r)\in\F_1$ denotes the pairings in which the momenta do
not participate in the change of variables and $(v_l,v_r)\in\F_2$
denotes the affected pairings. The explicit form of $z_l$ that appears
above is not important, so we do not specify them. The bounds
(\ref{apr3010}) and (\ref{apr3012}) mean that the ``participating''
$w_l$ give us the factor
\[
\prod_{(v_l,v_r)\in\F_2}|\hat{\phi}_0|(w_l)
\]
that appears in the last line of (\ref{eq:mbcv}).

Next, we integrate in time. This brings about the product
\[
C^k\prod_{(v_l,v_r)\in\F_1} \frac{\hat R(w_l)}{\g(w_l)}\prod_{(v_l,v_r)\in\F_2}\frac{\hat R(z_l)}{\g(z_l)}.
\]
Using the fact that $\hat{R}(w_l)/\g(w_l)$ is integrable for the
vertices in ${\cal F}_1$, and that $\hat{R}(z_l)/\g(z_l)$ is uniformly
bounded for the vertices in ${\cal F}_2$, we may integrate out all the
momenta variables, showing that \eqref{eq:mmbd3} is bounded by
\begin{equation}
\frac{\#(\F)}{\prod_{i=1}^M (m_i)!\prod_{j=1}^N (n_j)!}C^k=
\frac{(2k-1)!!}{\prod_{i=1}^M (m_i)!\prod_{j=1}^N (n_j)!}C^k
\end{equation}
This finishes the proof.~$\Box$

Lemma~\ref{lem:bdmm} ensures we can interchange the limit $\eps\to 0$
and the summation, since 
\[
 \sum_{m_1,\ldots,m_M=0}^\infty\sum_{n_1,\ldots,n_N=0}^\infty\frac{(2k-1)!!}{\prod_{i=1}^M (m_i)!\prod_{j=1}^N (n_j)!}C^k=\sum_{k=0}^\infty \frac{(M+N)^{2k}C^k}{2^kk!}<\infty.
\]

\subsubsection*{An estimate on non-simple pairings}

Now we need to consider more carefully the contribution from different
types of pairings.  First we can decompose the temporal domain
$\Delta_{m_1}(t)\times\ldots\times\Delta _{n_N}(t)$ according to all
possible permutations of~$\{s_{1,1},\ldots u_{N,n_N}\}$ and write
\begin{equation}
\begin{aligned}
&\E\{g_{m_1,\eps}\ldots g_{m_M,\eps} g_{n_1,\eps}^*\ldots g_{n_N,\eps}^*\}\\
=&\sum_{\sigma}\frac{1}{(i\eps)^{\sum_{i=1}^Mm_i}}
\frac{1}{(-i\eps)^{\sum_{j=1}^Nn_j}}\int_{\sigma_{2k}(t)} dsdu
\int_{\R^{2kd}} 
\E\{I_{M,N}\} e^{i\G_M}e^{-i\G_N}\prod_{i=1}^M  h_{M,i}\prod_{j=1}^N h_{N,j}^*,
\label{eq:defPer}
\end{aligned}
\end{equation}
where $\sigma_{2k}(t)=\{0\leq v_{2k}\leq \ldots\leq v_1\leq t\}$ and
$\sigma=\{v_1,\ldots,v_{2k}\}$ denotes all possible permutations of
$\{s_{1,1},\ldots u_{N,n_N}\}$ such that
$\sigma_{2k}(t)\neq \varnothing$. By \eqref{eq:IMN},
\begin{equation}
\E\{I_{M,N}\}=\sum_{\F} \prod_{(v_l,v_r)\in\F}(2\pi)^{-d}
e^{-\g(w_l)|v_{l}-v_{r}|/\eps^2}
\delta(w_l+ w_r)\hat{R}(w_l)dw_ldw_r,
\end{equation}
where $\F$ are pairings obtained by computing joint moments of
Gaussian. 
We can write 
\begin{equation}
\E\{g_{m_1,\eps}\ldots g_{m_M,\eps} g_{n_1,\eps}^*\ldots
g_{n_N,\eps}^*\}
=\sum_{\sigma}\sum_{\F} J_{m_1,\ldots,n_N^*}^\eps(\sigma,\F,\xi,g)
\end{equation}
with
\begin{equation}
\begin{aligned}
&J_{m_1,\ldots,n_N^*}^\eps(\sigma,\F,\xi,g)=
\frac{1}{(i\eps)^{\sum_{i=1}^Mm_i}}\frac{1}{(-i\eps)^{\sum_{j=1}^Nn_j}}\\
\times &\int_{\sigma_{2k}(t)} dsdu \int_{\R^{2kd}} 
\prod_{(v_l,v_r)\in\F}(2\pi)^{-d} e^{-\g(w_l)|v_{l}-v_{r}|/\eps^2}
\delta(w_l+ w_r)\hat{R}(w_l)dw_ldw_r e^{i\G_M}e^{-i\G_N}
\prod_{i=1}^M  h_{M,i}\prod_{j=1}^N h_{N,j}^*
\label{eq:defJ}
\end{aligned}
\end{equation}
and the symbol $g=f$ or $F$ indicates the dependence of
$J_{m_1,\ldots,n_N^*}^\eps$ on $g_{n,\eps}=f_{n,\eps}$ or
$F_{n,\eps}$.

Given a permutation $\sigma$, we say that $\F_\sigma$ is a
\emph{simple pairing} if $v_{2i-1},v_{2i}$ form a pair for every index
$i=1,\ldots,k$. The next lemma shows that the overall contribution of 
the non-simple pairings vanishes in the
limit $\eps\to 0$.
\begin{lemma}
Let $g_{n,\eps}=f_{n,\eps}$ or $F_{n,\eps}$, then we have 
\[
\sum_\sigma \sum_{\F\neq
  \F_\sigma}J_{m_1,\ldots,n_N^*}^\eps(\sigma,\F,\xi,g)\to 0,\hbox{ as
  $\eps\to 0$.}
\]
\label{lem:nonsimple}
\end{lemma}
{\bf Proof.}
When $g_{n,\eps}=f_{n,\eps}$, this is proved in \cite[Lemma
3.6]{BKR-ARMA-11}. The proof for $g_{n,\eps}=F_{n,\eps}$ is similar, using the
same change of variables as in the proof of Lemma~\ref{lem:bdmm}. We do not
provide all details -- but just mention the main simple point: for $\F\neq \F_\sigma$, i.e., the non-simple pairings,
the non-consecutive times are paired, then the overall contribution from the time integration of the
exponentials $e^{-\g(w_l)|v_{l}-v_{r}|/\eps^2}$ is too small since $|v_l-v_r|$ is too large.
We write
\begin{equation*}
\begin{aligned}
&|J_{m_1,\ldots,n_N^*}^\eps(\sigma,\F,\xi,F) |\\
\leq & \frac{1}{\eps^{2k}}\int_{\sigma_{2k}(t)} dsdu \int_{\R^{2kd}} 
\prod_{(v_l,v_r)\in\F}(2\pi)^{-d} e^{-\g(w_l)|v_{l}-v_{r}|/\eps^2}
\delta(w_l+ w_r)\hat{R}(w_l)dw_ldw_r\prod_{i=1}^M |h_{M,i}|\prod_{j=1}^N |h_{N,j}^*|,
\end{aligned}
\end{equation*}
and by the proof of \eqref{eq:mbcv}, we have
\begin{equation*}
\begin{aligned}
&|J_{m_1,\ldots,n_N^*}^\eps(\sigma,\F,\xi,F) |\\
\leq & \frac{C^{M+N}}{\eps^{2k}}\int_{\sigma_{2k}(t)} dsdu
\int_{\R^{kd}} 
\prod_{(v_l,v_r)\in\F_1}(2\pi)^{-d}
e^{-\g(w_l)|v_{l}-v_{r}|/\eps^2}\hat{R}(w_l)
\prod_{(v_l,v_r)\in\F_2}(2\pi)^{-d} e^{-\g(\cdot)|v_{l}-v_{r}|/\eps^2}\hat{R}(\cdot)\\
 &\times\prod_{(v_l,v_r)\in\F_2}|\hat{\phi}_0|(w_l)dw.
\end{aligned}
\end{equation*}
Then, using the fact that $\hat{R}(p)/\g(p)$ is integrable and
uniformly bounded, we only need to follow the proof of \cite[Lemma
3.6]{BKR-ARMA-11} using the aforementioned observation that the time
integration will bring about too high power of $\eps$ because of the
exponential in time factors.~$\Box$

\subsubsection*{The vanishing of the crossing pairings}

By Lemma~\ref{lem:nonsimple}, we have
\begin{equation}
\lim_{\eps\to 0} \E\{g_{m_1,\eps}\ldots g_{m_M,\eps} g_{n_1,\eps}^*\ldots g_{n_N,\eps}^*\}=\sum_\sigma \lim_{\eps\to 0}J_{m_1,\ldots,n_N^*}^\eps(\sigma,\F_\sigma,\xi,g).
\end{equation}
Let us define sets 
\[
\hbox{$A_i=\{s_{i,1},\ldots,s_{i,m_i}\}$, $B_j=\{u_{j,1},\ldots,u_{j,n_j}\}$ with $i=1,\ldots,M$, $j=1,\ldots,N$. }
\]
Given a pairing $\F_\sigma$, we say 
\[
S_1,S_2\in \{A_i,B_j: i=1,\ldots,M,j=1,\ldots,N\}
\]
interact with each other 
if there is an edge $(v_l,v_r)\in \F_\sigma$ such that $v_l\in S_1,v_r\in S_2$, and we write~$S_1\leftrightarrow S_2$. 
We say they are connected if there exist other sets such that $S_1\leftrightarrow\ldots \leftrightarrow S_2$.
Thus, for a given permutation $\sigma$, we may decompose $\{A_i,B_j: i=1,\ldots,M,j=1,\ldots,N\}$ into connected components. 
For example, if all variables in $A_1$ pair inside $A_1$, then $A_1$ itself is a connected component.
If all variables in $A_1$ and $A_2$ either pair inside the corresponding set or pair with variables in the other set, 
and we have at least one edge joining $A_1$ and $A_2$, then~$\{A_1,A_2\}$ is a connected component, and so on. 
We let $N_c(\F_\sigma)$ be the size of largest connected component corresponding to $\F_\sigma$. 
The following lemma shows the permutations with more than triple interactions do not contribute in the limit. This leads to 
a Gaussian limit in Theorems~\ref{thm:trHigh} and \ref{thm:cor}.
\begin{lemma} We have
\[
\sum_{\sigma: N_c(\F_\sigma)\geq 2} \lim_{\eps\to 0}J_{m_1,\ldots,n_N^*}^\eps(\sigma,\F_\sigma,\xi,f)=0.
\]
and 
\[
\sum_{\sigma: N_c(\F_\sigma)\geq 3} \lim_{\eps\to 0}J_{m_1,\ldots,n_N^*}^\eps(\sigma,\F_\sigma,\xi,F)=0.
\]
\label{lem:nontriple}
\end{lemma}
{\bf Proof.}
We first consider the case $g_{n,\eps}=f_{n,\eps}$. For a given permutation $\sigma$, if $N_c(\sigma)\geq 2$, 
we can find the sets
\[
S_1,S_2\in\{A_i,B_j: i=1,\ldots,M,j=1,\ldots,N\},
\]
 such that $S_1\leftrightarrow S_2$. Let $e$ be an edge joining $S_1$ and $S_2$, and $h_{S_1},h_{S_2}$ 
 be the initial conditions corresponding to $S_1,S_2$, then we have
\begin{equation}
\begin{aligned}
&|J_{m_1,\ldots,n_N^*}^\eps(\sigma,\F_\sigma,\xi,f) |\\
\leq & \frac{C^{M+N-2}}{\eps^{2k}}\int_{\sigma_{2k}(t)} dsdu \int_{\R^{2kd}} 
\prod_{(v_l,v_r)\in\F_\sigma}(2\pi)^{-d} e^{-\g(w_l)|v_{l}-v_{r}|/\eps^2}\delta(w_l+ w_r)\hat{R}(w_l)dw_ldw_r|h_{S_1}h_{S_2}|,
\label{eq:nontri1}
\end{aligned}
\end{equation}
where other factors of the initial condition $\hat{\phi}_0$ are bounded by $C^{M+N-2}$. Recall that when $g_{n,\eps}=f_{n,\eps}$, we have
\[
h_{M,i}=\hat{\phi}_0(\xi-\frac{p_{i,1}+\ldots+p_{i,m_i}}{\eps^\alpha}),
~~h_{N,j}^*=\hat{\phi}_0^*(\xi-\frac{q_{j,1}+\ldots+q_{j,n_j}}{\eps^\alpha}).
\]
We can assume 
\[
|h_{S_1}|=|\hat{\phi}_0|(\xi-\frac{P_1}{\eps^\alpha})\hbox{ and }|h_{S_2}|=|\hat{\phi}_0|(\xi-\frac{P_2}{\eps^\alpha}),
\]
for some $P_1,P_2$ after integrating out $w_r$ in \eqref{eq:nontri1}. 
It is clear that $P_1,P_2\neq 0$ since they both contain the $w_l$ variable corresponding to the edge $e$. Now we integrate $w_r$ and the time variables to obtain
\begin{equation}
|J_{m_1,\ldots,n_N^*}^\eps(\sigma,\F_\sigma,\xi,f) |\leq  C^{M+N}\int_{\R^{kd}}\prod_{(v_l,v_r)\in\F_\sigma} 
\frac{\hat{R}(w_l)}{\g(w_l)}|\hat{\phi}_0|(\xi-\frac{P_1}{\eps^\alpha})|\hat{\phi}_0|(\xi-\frac{P_2}{\eps^\alpha})|dw_l\to 0,
\end{equation}
as $\eps\to 0$ by dominated convergence theorem.

Next we consider the case $g_{n,\eps}=F_{n,\eps}$. The following estimate holds 
\begin{equation*}
\begin{aligned}
&|J_{m_1,\ldots,n_N^*}^\eps(\sigma,\F_\sigma,\xi,F) |\\
\leq & \frac{1}{\eps^{2k}}\int_{\sigma_{2k}(t)} dsdu \int_{\R^{2kd}} \prod_{(v_l,v_r)\in\F_\sigma}(2\pi)^{-d} e^{-\g(w_l)|v_{l}-v_{r}|/\eps^2}\delta(w_l+ w_r)\hat{R}(w_l)dw_ldw_r\prod_{i=1}^M |h_{M,i}|\prod_{j=1}^N |h_{N,j}^*|.
\end{aligned}
\end{equation*}
Recall that now
\[
h_{M,i}=\eps^{-\alpha d/2}\hat{\phi}_0(\frac{\xi-p_{i,1}-\ldots-p_{i,m_i}}{\eps^\alpha}), 
~~~h_{N,j}^*=\eps^{-\alpha d/2}\hat{\phi}_0^*(\frac{\xi-q_{j,1}-\ldots-q_{j,n_j}}{\eps^\alpha}).
\]
If $N_c(\sigma)\geq 3$, we can find 
\[
S_1,S_2,S_3\in\{A_i,B_j: i=1,\ldots,M,j=1,\ldots,N\}
\]
such that $S_1\leftrightarrow S_2\leftrightarrow S_3$. We pick two edges linking $S_1$ to $S_2$ and $S_2$ to $S_3$, 
and denote them by~$e_{1,2}$ and $e_{2,3}$, respectively. We also denote the variables corresponding to $e_{1,2},e_{2,3}$ by 
$w_{1,2},w_{2,3}$. Let $h_{S_i}$ be the initial condition corresponding to $S_i,i=1,2,3$, then we have 
\[
|h_{S_i}|=\eps^{-\alpha d/2}|\hat{\phi}_0|(\frac{\xi-P_i}{\eps^\alpha})\hbox{ for some $P_i,i=1,2,3$. }
\]
After integrating out the $w_r$ variables, it is clear that $P_1$
contains the variable $w_{1,2}$, $P_2$ contains the variables
$w_{1,2},w_{2,3}$ and $P_3$ contains the variable $w_{2,3}$.  We do a
similar change of variable as in the proof of Lemma~\ref{lem:bdmm}.
First, we change $w_{1,2}$ so that
${(\xi-P_1)}/{\eps^\alpha}\mapsto P_1$. Second, we change $w_{2,3}$ so
that $(\xi-P_3)/{\eps^\alpha}\mapsto P_3$. Then we have
\begin{equation}
\begin{aligned}
|h_{S_1}h_{S_2}h_{S_3}|=&\eps^{-3\alpha d/2}|\hat{\phi}_0(\frac{\xi-P_1}{\eps^\alpha})
\hat{\phi}_0(\frac{\xi-P_2}{\eps^\alpha})\hat{\phi}_0(\frac{\xi-P_3}{\eps^\alpha})|dw_{1,2}dw_{2,3}\\
\mapsto &\eps^{-3\alpha d/2}\eps^{2\alpha d}|\hat{\phi}_0(P_1)\hat{\phi}_0(z)\hat{\phi}_0(P_3)|dP_1dP_3
\leq C\eps^{\alpha d/2}|\hat{\phi}_0(P_1)\hat{\phi}_0(P_3)|dP_1dP_3,
\end{aligned}
\end{equation}
with some $z$ that does not matter to us, as we simply bound
$\hat{\phi}_0(z)$ by $C$.  Now, we only need to carry out the same
change of variable as in the proof of Lemma~\ref{lem:bdmm} for the
remaining $h$. In the end, we obtain
\begin{equation}
\begin{aligned}
&|J_{m_1,\ldots,n_N^*}^\eps(\sigma,\F_\sigma,\xi,F) |
\leq  \frac{C^{M+N}\eps^{\alpha d/2}}{\eps^{2k}}\int_{[0,t]^{2k}} dsdu \int_{\R^{kd}} 
\prod_{(v_l,v_r)\in\F_{\sigma,1}}(2\pi)^{-d} e^{-\g(w_l)|v_{l}-v_{r}|/\eps^2}\hat{R}(w_l)\\
&\times\prod_{(v_l,v_r)\in\F_{\sigma,2}}(2\pi)^{-d} e^{-\g(z_l)|v_{l}-v_{r}|/\eps^2}\hat{R}(z_l)
\prod_{(v_l,v_r)\in\F_{\sigma,2}}|\hat{\phi}_0|(w_l)dw,
 \label{eq:nontri2}
\end{aligned}
\end{equation}
where $(v_l,v_r)\in\F_{\sigma,1}$ denotes the pairings which are not
affected by the change of variables, and~$(v_l,v_r)\in\F_{\sigma,2}$
denotes the affected pairings, and, as in the analysis of
(\ref{eq:mbcv}), the precise expression for $z_l$ is not important to
us. Clearly, the RHS of \eqref{eq:nontri2} goes to zero as $\eps\to 0$
because of the extra factor $\eps^{\alpha d/2}$ compared to
(\ref{eq:mbcv}).~$\Box$

\subsubsection*{Pairings for the correctors}

We now describe analogous estimates that are needed in the analysis of the corrector
\[
\U_\eps(t,\xi)=\eps^{-\alpha d/2}\sum_{n=0}^\infty \FF_{n,\eps}(t,\xi),
\]
with
\begin{equation*}
\begin{aligned}
&\!\!\FF_{n,\eps}(t,\xi)=f_{n,\eps}(t,\xi)-\E\{f_{n,\eps}(t,\xi)\}\\
&~~~~~~~~~~~=\frac{1}{(i\eps)^n}\int_{\Delta_n(t)}\int_{\R^{nd}}
\!\V(\frac{s_1}{\eps^2},\ldots,\frac{s_n}{\eps^2},dp_1,\ldots,dp_n)
e^{iG_n(\eps^\alpha \xi, s^{(n)},p^{(n)})/\eps^{2}} \!\hat{\phi}_0(\xi-\frac{p_1}{\eps^\alpha}-\ldots-\frac{p_n}{\eps^\alpha})ds,
\end{aligned}
\end{equation*}
where 
\begin{equation}\label{apr3020}
\V(s_1,\ldots,s_n,dp_1,\ldots,dp_n)=(2\pi)^{-nd}\left(\prod_{j=1}^n \tilde{V}(s_j,dp_j)-\E\{\prod_{j=1}^n \tilde{V}(s_j,dp_j )\}\right).
\end{equation}
Let us discuss in detail the calculation of moments
\[
\eps^{-\alpha d(M+N)/2}\E\{\FF_{m_1,\eps}\ldots\FF_{m_M,\eps}\FF_{n_1,\eps}^*\ldots\FF_{n_N,\eps}^*\},
\hbox{ for $m_1,\ldots,n_N\in\mathbb{N}$,}
\]
with 
\[
\sum_{i=1}^M m_i+\sum_{j=1}^N n_j=2k.
\]
Similar to \eqref{eq:mm}, we can write
\begin{equation}
\begin{aligned}
&\eps^{-\alpha d(M+N)/2}\E\{\FF_{m_1,\eps}\ldots\FF_{m_M,\eps}\FF_{n_1,\eps}^*\ldots\FF_{n_N,\eps}^*\}\\
=&\frac{1}{(i\eps)^{\sum_{i=1}^Mm_i}}\frac{1}{(-i\eps)^{\sum_{j=1}^Nn_j}}
\int_{\Delta_{m_1}(t)\times\ldots\times\Delta _{n_N}(t)} dsdu \int_{\R^{2kd}} 
\E\{\I_{M,N}\} e^{i\G_M}e^{-i\G_N} \prod_{i=1}^M  h_{M,i}\prod_{j=1}^N h_{N,j}^*,
\label{eq:mmcor}
\end{aligned}
\end{equation}
where 
\begin{equation}
\begin{aligned}
\I_{M,N}=&\V(\frac{s_{1,1}}{\eps^2},\ldots,\frac{s_{1,m_1}}{\eps^2},dp_{1,1},\ldots,dp_{1,m_1})\ldots \V(\frac{s_{M,1}}{\eps^2},\ldots,\frac{s_{M,m_M}}{\eps^2},dp_{M,1},\ldots,dp_{M,m_M})\\
&\times \V^*(\frac{u_{1,1}}{\eps^2},\ldots,\frac{u_{1,n_1}}{\eps^2},dq_{1,1},\ldots,dq_{1,n_1})\ldots \V^*(\frac{u_{N,1}}{\eps^2},\ldots,\frac{u_{N,n_N}}{\eps^2},dq_{N,1},\ldots,dq_{N,n_N}),
\end{aligned}
\end{equation}
and 
\[
h_{M,i}=\eps^{-\alpha d/2}\hat{\phi}_0(\xi-\frac{p_{i,1}+\ldots+p_{i,m_i}}{\eps^\alpha}),
~~~h_{N,j}^*=\eps^{-\alpha d/2}\hat{\phi}_0^*(\xi-\frac{q_{j,1}+\ldots+q_{j,n_j}}{\eps^\alpha}).
\]
Previously, we have dealt with the expectation of a product of
centered Gaussians. For $\I_{M,N}$, however, each factor $\V$, defined
in (\ref{apr3020}) is a centered product of Gaussians rather than a
product of centered Gaussians. The rules for evaluating the expectation 
of such objects are recalled in Lemma~\ref{lem:nonself} in the Appendix.
Recall that we have defined the sets 
\[
A_i=\{s_{i,1},\ldots,s_{i,m_i}\},~~
B_j=\{u_{j,1},\ldots,u_{j,n_j}\}, \hbox{ with $i=1,\ldots,M$,
$j=1,\ldots,N$.}
\]
Given a pairing $\F$, we decompose $\{A_i,B_j:
i=1,\ldots,M,j=1,\ldots,N\}$ into connected components according to the
interaction between the $s,u$ variables. Let $N_s(\F)$ be the size of
smallest connected component, then by Lemma~\ref{lem:nonself} we have
\begin{equation}\label{may102}
\E\{\I_{M,N}\}=\sum_{\F:N_s(\F)\geq 2} 
\prod_{(v_l,v_r)\in\F}(2\pi)^{-d} e^{-\g(w_l)|v_{l}-v_{r}|/\eps^2}\delta(w_l+ w_r)\hat{R}(w_l)dw_ldw_r.
\end{equation}
In particular, it is clear that $\E\{\I_{M,N}\}\leq \E\{I_{M,N}\}$ and 
\begin{equation}
\E\{I_{M,N}\}-\E\{\I_{M,N}\}=
\sum_{\F:N_s(\F)=1} \prod_{(v_l,v_r)\in\F}(2\pi)^{-d} e^{-\g(w_l)|v_{l}-v_{r}|/\eps^2}\delta(w_l+ w_r)\hat{R}(w_l)dw_ldw_r.
\end{equation}
Comparing \eqref{eq:mm} and (\ref{eq:IMN}), to (\ref{eq:mmcor}) and
(\ref{may102}), 
we see that 
\[
\eps^{-\alpha d(M+N)/2}\E\{\FF_{m_1,\eps}\ldots\FF_{m_M,\eps}\FF_{n_1,\eps}^*\ldots\FF_{n_N,\eps}^*\}.
\]
has exactly the same form as
\begin{equation}\label{may104}
\E\{F_{m_1,\eps}\ldots F_{m_M,\eps}F_{n_1,\eps}^*\ldots F_{n_N,\eps}^*\}
\end{equation}
if we replace $\xi\to \eps^\alpha \xi$ and impose the constraint
$N_s(\F)\geq 2$ in (\ref{may104}).  Therefore, we can follow the same
proof for Lemmas~\ref{lem:bdmm}, \ref{lem:nonsimple} and obtain
\begin{equation*}
\lim_{\eps\to 0}\eps^{-\alpha d(M+N)/2}\E\{\FF_{m_1,\eps}\ldots\FF_{m_M,\eps}\FF_{n_1,\eps}^*\ldots\FF_{n_N,\eps}^*\}=\sum_{\sigma:N_s(\F_\sigma)\geq 2} \lim_{\eps\to 0}J_{m_1,\ldots,n_N^*}^\eps(\sigma,\F_\sigma,\eps^\alpha \xi,F),
\end{equation*}
where we recall $J_{m_1,\ldots,n_N^*}^\eps$ is defined in \eqref{eq:defJ}.

We should note that in the proof of Lemma~\ref{lem:bdmm}, for
$g_{n,\eps}=F_{n,\eps}$, we used the fact that $\xi\neq 0$ so that
\[
\eps^{-\alpha d/2}|\hat{\phi}_0(\xi/\eps^\alpha)|\leq C,
\]
and actually
goes to zero as $\eps\to 0$. At this step, the analysis 
for $\FF_{n,\eps}$ can not proceed this way, as we have replaced 
$\xi\mapsto \eps^\alpha \xi$. Instead, we use
the condition $N_s(\F)\geq 2$, which implies that after computing
moments, all the $h$ factors in \eqref{eq:mmcor} take the form
\[
h_{M,i}=\eps^{-\alpha d/2}\hat{\phi}_0\big(\xi-\frac{P}{\eps^\alpha}\big),
\hbox{ and }
h_{N,j}^*=\eps^{-\alpha d/2}\hat{\phi}_0\big(\xi-\frac{Q}{\eps^\alpha}\big),
\]
for some $P,Q\neq 0$. If $P$ or $Q$ were to be zero, then $A_i$ or
$B_j$ is not connected with any other set, which would imply
$N_s(\F)=1$. As $P$ and $Q$ are not zero, we only need to perform the same change of
variables as in the proof of Lemma~\ref{lem:bdmm}.

We may now follow the same proof as for Lemma~\ref{lem:nontriple} to obtain
\begin{equation}\label{may106}
\lim_{\eps\to 0}\eps^{-\alpha d(M+N)/2}\E\{\FF_{m_1,\eps}\ldots
\FF_{m_M,\eps}\FF_{n_1,\eps}^*\ldots\FF_{n_N,\eps}^*\}
\!=\!\!\!\!
\sum_{\sigma:N_s(\F_\sigma)\geq 2, N_c(\F_\sigma)\leq 2}\!\! 
\lim_{\eps\to 0}J_{m_1,\ldots,n_N^*}^\eps(\sigma,\F_\sigma,\eps^\alpha \xi,F).
\end{equation}
Since $N_s(\F_\sigma)\geq 2$ and $N_c(\F_\sigma)\leq 2$, we have
\[
N_s(\F_\sigma)=N_c(\F_\sigma)=2,
\]
that is, all connected components corresponding to $\F_\sigma$ contain
two sets, which implies $M+N$ is even.

\section{Homogenization of the low frequencies}\label{sec:homlow}

We now prove Theorem~\ref{thm:homo}.  To show that
\[
\psi_\eps(t,\xi)\to \hat{\phi}_0(\xi)e^{-\frac12D(0)t}\hbox{ in
probability,}
\]
we only need to verify the following result.
\begin{proposition}\label{prop:con1M}
 As $\eps\to 0$, we have
\begin{equation}\label{may110}
\E\{\psi_\eps(t,\xi)\}\to \hat{\phi}_0(\xi)e^{-\frac12D(0)t},
\end{equation}
and
\begin{equation}\label{may108}
\E\{|\psi_\eps(t,\xi)|^2\}\to |\hat{\phi}_0(\xi)|^2e^{-\red(0) t}.
\end{equation}
\end{proposition}
{\bf Proof.} 
By Lemma~\ref{lem:gnbd}, we have 
\[
\E\{\psi_\eps(t,\xi)\}=\sum_{n=0}^\infty \E\{f_{n,\eps}(t,\xi)\}.
\]
Lemma~\ref{lem:bdmm} ensures that we only need to compute
\[
\lim_{\eps\to 0} \E\{f_{n,\eps}(t,\xi)\},
\]
when $n=2k$ for some
$k\in\mathbb{N}$. By Lemma~\ref{lem:nonsimple}, we have
\begin{equation}
\lim_{\eps\to 0}\E\{f_{n,\eps}(t,\xi)\}=\sum_\sigma\lim_{\eps\to 0}J_n^\eps(\sigma,\F_\sigma,\xi,f).
\end{equation}
It is straightforward to see that
\begin{equation}
\begin{aligned}
&J_n^\eps(\sigma,\F_\sigma,\xi,f)\\
=&\frac{\hat{\phi}_0(\xi)}{(i\eps)^{2k}}\int_{\sigma_{2k}(t)} ds
\int_{\R^{kd}} 
\prod_{(v_l,v_r)\in\F_\sigma}(2\pi)^{-d}
e^{-\g(w_l)|v_{l}-v_{r}|/\eps^2}\hat{R}(w_l)
e^{i(|\eps^\alpha \xi|^2-|\eps^\alpha \xi-w_l|^2)\frac{|v_l-v_r|}{2\eps^2}}dw\\
\to & \hat{\phi}_0(\xi)(-1)^k\frac{t^k}{k!}\left(\int_{\R^d}
\frac{\hat{R}(p)}{(\g(p)+\frac{i}{2}|p|^2)}\farc{dp}{(2\pi)^d}\right)^k
=\hat{\phi}_0(\xi)\frac{(-tD(0)/2)^k}{k!},
\label{eq:con1M1}
\end{aligned}
\end{equation}
and thus 
\[
\lim_{\eps\to 0}\E\{\psi_\eps(t,\xi)\}=\sum_{k=0}^\infty 
\lim_{\eps\to 0}\E\{f_{2k,\eps}(t,\xi)\}=\sum_{k=0}^\infty 
\lim_{\eps\to 0} J_{2k}^\eps(\sigma,\F_\sigma,\xi,f)=\hat{\phi}_0(\xi)e^{-\frac12D(0)t},
\]
which is (\ref{may110}).
 
Since 
\[
\E\{|\psi_\eps(t,\xi)|^2\}=\sum_{m,n=0}^\infty
\E\{f_{m,\eps}(t,\xi)f_{n,\eps}^*(t,\xi)\},
\]
by a similar discussion as in the proof of  (\ref{may110}), we have 
\begin{equation}
\lim_{\eps\to 0}\E\{f_{m,\eps}(t,\xi)f_{n,\eps}^*(t,\xi)\}=\sum_\sigma \lim_{\eps\to 0}J_{m,n^*}^\eps(\sigma,\F_\sigma,\xi,f)
\end{equation}
In addition, Lemma~\ref{lem:nontriple} shows that 
\begin{equation}
\sum_{\sigma:N_c(\sigma)\geq 2} \lim_{\eps\to 0}J_{m,n^*}^\eps(\sigma,\F_\sigma,\xi,f)=0,
\end{equation} 
so we are left with
\[
\sum_{\sigma:N_c(\sigma)=1}\lim_{\eps\to 0}J_{m,n^*}^\eps(\sigma,\F_\sigma,\xi,f).
\]
However, $N_c(\sigma)=1$ implies there is no interaction between
$f_{m,\eps}(t,\xi)$ and $f_{n,\eps}^*(t,\xi)$, so $m=2k_1$, and~$n=2k_2$ are
both even. The number of possible permutations is
\[
\frac{(k_1+k_2)!}{k_1!k_2!},
\]
and by the same calculation for
\eqref{eq:con1M1}, we have
\begin{equation}
\begin{aligned}
\sum_{\sigma:N_c(\sigma)=1}\lim_{\eps\to
  0}J_{m,n^*}^\eps(\sigma,\F_\sigma,\xi,f)=
&\frac{(k_1+k_2)!}{k_1!k_2!} |\hat{\phi}_0(\xi)|^2(-1)^{k_1+k_2}
\frac{t^{k_1+k_2}}{(k_1+k_2)!}\left(\frac{D(0)}{2}\right)^{k_1}\left(\frac{D^*(0)}{2}\right)^{k_2}\\
=&|\hat{\phi}_0(\xi)|^2\frac{(-tD(0)/2)^{k_1}}{k_1!}\frac{(-tD^*(0)/2)^{k_2}}{k_2!}
\end{aligned}
\end{equation}
Therefore, we have
\begin{equation}
\lim_{\eps\to 0}\E\{|\psi_\eps(t,\xi)|^2\}=\sum_{k_1,k_2=0}^\infty |\hat{\phi}_0(\xi)|^2\frac{(-tD(0)/2)^{k_1}}{k_1!}\frac{(-tD^*(0)/2)^{k_2}}{k_2!}=|\hat{\phi}_0(\xi)|^2e^{-\red(0)t},
\end{equation}
which is (\ref{may108}).~$\Box$

\section{The high frequencies}\label{sec:high}

In this section, we prove Theorem~\ref{thm:trHigh}.

\subsection*{Convergence of the mean}

We first show the convergence of 
$\E\{\Psi_\eps(t,\xi)\}$ for fixed $t>0$ and $\xi\neq 0$.
\begin{lemma}\label{lem:con1mTr} 
We have
\[
\E\{\Psi_\eps(t,\xi)\}\to 0\hbox{ as $\eps\to 0$.}
\]
\end{lemma}
{\bf Proof.}
By Lemmas~\ref{lem:bdmm}, \ref{lem:nonsimple}, 
we only need to show that
\[
\lim_{\eps\to 0}J_n^\eps(\sigma,\F_\sigma,\xi,F)=0,
\]
when $n=2k$. It is straightforward to see that
\begin{eqnarray*}
&&J_n^\eps(\sigma,\F_\sigma,\xi,F)=\eps^{-\alpha d/2}\hat{\phi}_0(\frac{\xi}{\eps^\alpha})\\
&&\times\frac{1}{(i\eps)^{2k}}\int_{\sigma_{2k}(t)} ds \int_{\R^{kd}} 
\prod_{(v_l,v_r)\in\F_\sigma}(2\pi)^{-d}
   e^{-\g(w_l)|v_{l}-v_{r}|/\eps^2}
\hat{R}(w_l)e^{i(|\eps^\alpha \xi|^2-|\eps^\alpha \xi-w_l|^2)\frac{|v_l-v_r|}{2\eps^2}}dw.
\end{eqnarray*}
Since $\xi\neq 0$, we have
\[
\eps^{-\alpha d/2}\hat{\phi}_0(\xi/\eps^\alpha)\to 0\hbox{ as $\eps\to
  0$,}
\]
thus
\[
|J_n^\eps(\sigma,\F_\sigma,\xi,F)|\leq C\eps^{-\alpha
  d/2}\hat{\phi}_0(\xi/\eps)\to 0\hbox{ as $\eps\to 0$},
\]
and the proof is complete.~$\Box$

\subsection*{Convergence of the variance}

Next, we look at the second moment.
\begin{lemma}\label{lem:con2mTr}
We have
\[
\E\{|\Psi_\eps(t,\xi)|^2\}\to \widehat{W}_\delta(t,\xi)\hbox{ as
  $\eps\to 0$.}
\]
\end{lemma}
The proof of Lemma~\ref{lem:con2mTr} is very similar to
\cite[Proposition 3.12]{BKR-ARMA-11}, and since Lemmas~\ref{lem:corab2m} and
\ref{lem:cor2m} below follow the same blueprint, we will provide the
details here for the convenience of the reader.

{\bf Proof}. By Lemmas~\ref{lem:gnbd}-\ref{lem:nontriple},
we only need to consider $J_{m,n^*}^\eps(\sigma,\F_\sigma,\xi,F)$ for
fixed $m,n\in\mathbb{N}$ (in the present case, we automatically have
$N_c(\F_\sigma)\leq 2$). We write 
\[
A=\{s_1,\ldots,s_m\},~~~
B=\{u_1,\ldots,u_n\},
\]
with $m+n=2k$ for some
$k\in\mathbb{N}$. According to the pairing $\F_\sigma$, $\{A,B\}$ is decomposed into
connected components. If $A,B$ are ``separate", we have two factors of
$\eps^{-\alpha d/2}|\hat{\phi}_0|(\xi/\eps^\alpha)$ coming from the
initial conditions, so by the same argument as in the proof of
Lemma~\ref{lem:con1mTr}, we have
\[
J_{m,n^*}^\eps(\sigma,\F_\sigma,\xi,F)\to 0\hbox{ as $\eps\to 0$.}
\]
Therefore, we only need to consider $\sigma$ such that
$A\leftrightarrow B$.

For a permutation $\sigma$ of $A\cup B$, the simple diagram
$\F_\sigma$ corresponds to
\[
A\cup B=\{v_1^+,v_1^-, \ldots,v_k^+,v_k^-\},
\]
with
\[
v_1^+\geq v_1^- \ldots\geq v_k^+\geq v_k^-,
\]
and $(v_i^+,v_i^-)$ forming a pair, $i=1,\ldots,k$. Since
$A\leftrightarrow B$, there exists at least one pair such that~$v_i^+$
and $v_i^-$ come from different sets, and we call such pair a crossing
edge between $A$ and~$B$.  Assuming the total number of crossing edges
is $N_{cr}(\sigma)\geq 1$, the interval~$[0,t]$ is decomposed into
$N_{cr}+1$ parts according to the position of those crossing edges,
which we denote by
\[
r_1^+\geq r_1^-\geq \ldots \geq r_{N_{cr}}^+\geq r_{N_{cr}}^-,
\]
with $r_i^\pm=v_j^\pm$ for some $j$, and with the convention where
$r_0^-=t$ and $r_{N_{cr}+1}^+=0$. We further denote by
$\mathcal{E}_{s,i}$, $i=0,\ldots,N_{cr}$, the set of edges between the
vertices~$r_i^-$ and~$r_{i+1}^+$ that are of the form~$(s_j,s_{j+1})$,
and by $\mathcal{E}_{u,i}$ the set of edges between~$r_i^-$
and~$r_{i+1}^+$ that are of the form $(u_j,u_{j+1})$. The
corresponding sets of indices are denoted by
\[
\mathcal{A}_i=\{j: (v_j^+,v_j^-)\in\mathcal{E}_{s,i},j=1,\ldots,k\},
\]
and
\[
\mathcal{B}_i=\{j: (v_j^+,v_j^-)\in\mathcal{E}_{u,i},j=1,\ldots,k\},
\]
with $i=0,\ldots,N_{cr}$. For a non-crossing edge $(v_j^+,v_j^-)$, we
denote $\tau_j=1$ if $v_j^+,v_j^-$ are $s-$variables and $\tau_j=-1$ if
they are $u-$variables.

Recall that
\begin{equation}
\begin{aligned}
&J_{m,n^*}^\eps(\sigma,\F_\sigma,\xi,F)\\
=&\frac{1}{(i\eps)^{m}(-i\eps)^n}\int_{\sigma_{2k}(t)}
dsdu\int_{\R^{2kd}} 
\prod_{(v_l,v_r)\in\F_\sigma}(2\pi)^{-d}
e^{-\g(w_l)|v_{l}-v_{r}|/\eps^2}\delta(w_l+ w_r)
\hat{R}(w_l)dw_ldw_r\\
&\times e^{iG_m( \xi, s^{(m)},p^{(m)})/\eps^{2}}e^{-iG_n( \xi,
  u^{(n)},q^{(n)})/\eps^{2}} 
\eps^{-\alpha d}\hat{\phi}_0(\frac{\xi-p_1-\ldots-p_m}{\eps^\alpha})
\hat{\phi}_0^*(\frac{\xi-q_1-\ldots-q_n}{\eps^\alpha}),
\label{eq:2mTrc1}
\end{aligned}
\end{equation}
where $v_l,v_r$ are the vertices of a given pair, and $w_l,w_r$ are
the corresponding $p,q$ variables, that is,~$w_l=p_i$ if $v_l=s_i$ and
$w_l=-q_i$ if $v_l=u_i$. For a crossing edge
$(v_l,v_r)=(r_i^+,r_i^-)$, the relevant~$p,q$
variables equal to each other due to $\delta(p-q)$, and we denote the
corresponding $w_l=P_i$, with the convention that $P_0=0$. We also define
$\mathfrak{s}_i=1$ if $r_i^+$ is $s-$variable and $\mathfrak{s}_i=-1$
if $r_i^+$ is $u-$variable.
With the above notation, we have
\begin{equation}
\begin{aligned}
&J_{m,n^*}^\eps(\sigma,\F_\sigma,\xi,F)\\
=&\frac{1}{(i\eps)^{m}(-i\eps)^n}\int_{\sigma_{2k}(t)}
dsdu\int_{\R^{kd}} dw
\prod_{j=1}^{k}\frac{\hat{R}(w_j)}{(2\pi)^d}e^{-\g(w_j)\frac{v_j^+-v_j^-}{\eps^2}}
\prod_{j=1}^{N_{cr}} e^{i\mathfrak{s}_j(|\xi-\ldots-P_{j-1}|^2-|\xi-\ldots-P_j|^2)\frac{r_j^+-r_j^-}{2\eps^2}}\\
&\times \prod_{j=0}^{N_{cr}}\prod_{l\in \mathcal{A}_j\cup \mathcal{B}_j}
e^{i\tau_l(|\xi-\ldots-P_j|^2-|\xi-\ldots-P_j-w_l|^2)\frac{v_l^+-v_l^-}{2\eps^2}} 
\frac{1}{\eps^{\alpha
    d}}|\hat{\phi}_0|(\frac{\xi-P_1-\ldots-P_{N_{cr}}}{\eps^\alpha})^2 .
\label{eq:2mTrc2}
\end{aligned}
\end{equation}
Here, we have integrated out the variables $w_r$ in \eqref{eq:2mTrc1}, 
and changed the notation $w_l\mapsto w_j$. 
To get rid of the extra factor ${\eps^{-\alpha d}}$, we change variables as before. Replacing 
\[
P_{N_{cr}}\mapsto \xi-P_1-\ldots-P_{N_{cr}-1}-\eps^\alpha P_{N_{cr}},
\] 
and rewriting the terms in \eqref{eq:2mTrc2} associated with $P_{N_{cr}}$ using the new variable, we obtain
\begin{equation*}
\begin{aligned}
&J_{m,n^*}^\eps(\sigma,\F_\sigma,\xi,F)
=\frac{1}{(i\eps)^{m}(-i\eps)^n}\int_{\sigma_{2k}(t)}
dsdu\int_{\R^{kd}} dw \prod_{j:w_j\neq P_{N_{cr}}}\frac{\hat{R}(w_j)}{(2\pi)^d}e^{-\g(w_j)\frac{v_j^+-v_j^-}{\eps^2}}\\
&\times 
\frac{\hat{R}(\xi-P_1-\ldots-P_{N_{cr}-1}-\eps^\alpha P_{N_{cr}})}{(2\pi)^d}
e^{-\g(\xi-P_1-\ldots-P_{N_{cr}-1}-\eps^\alpha P_{N_{cr}})\frac{r_{N_{cr}}^+-r_{N_{cr}}^-}{\eps^2}}\\
&\times \prod_{j=1}^{N_{cr}-1} 
\left(e^{i\mathfrak{s}_j(|\xi-\ldots-P_{j-1}|^2-|\xi-\ldots-P_j|^2)\frac{r_j^+-r_j^-}{2\eps^2}}\right)
e^{i\mathfrak{s}_{N_{cr}}(|\xi-\ldots-P_{N_{cr}-1}|^2-|\eps^\alpha P_{N_{cr}}|^2)\frac{r_{N_{cr}}^+-r_{N_{cr}}^-}{2\eps^2}}\\
&\times \prod_{j=0}^{N_{cr}-1}\left(\prod_{l\in \mathcal{A}_j\cup
    \mathcal{B}_j}
e^{i\tau_l(|\xi-\ldots-P_j|^2-|\xi-\ldots-P_j-w_l|^2)\frac{v_l^+-v_l^-}{2\eps^2}}\right)
\prod_{l\in \mathcal{A}_{N_{cr}}\cup\mathcal{B}_{N_{cr}}}
e^{i\tau_l(|\eps^\alpha P_{N_{cr}}|^2-|\eps^\alpha P_{N_{cr}}-w_l|^2)\frac{v_l^+-v_l^-}{2\eps^2}}\\
&\times |\hat{\phi}_0(P_{N_{cr}})|^2.
\end{aligned}
\end{equation*}
Now, we freeze $r_1^-\geq r_2^-\geq \ldots\geq r_{N_{cr}}^-$, 
integrate out the other time variables and send $\eps\to 0$ to obtain
\begin{equation*}
\begin{aligned}
&J_{m,n^*}^\eps(\sigma,\F_\sigma,\xi,F)\\
\to&(-1)^{n-k}\int_{\Delta_{N_{cr}}(t)} dv\int_{\R^{kd}} dw
\prod_{j=1}^{N_{cr}-1}\frac{1}{(2\pi)^d}
\frac{\hat{R}(P_j)}{\g(P_j)-i\mathfrak{s}_j(|\xi-\ldots-P_{j-1}|^2-|\xi-\ldots-P_j|^2)/2}\\
&\times \prod_{j=0}^{N_{cr}-1}
\prod_{l\in \mathcal{A}_j\cup \mathcal{B}_j}\frac{1}{(2\pi)^d}
\frac{\hat{R}(w_l)}{\g(w_l)-i\tau_l(|\xi-\ldots-P_j|^2-|\xi-\ldots-P_j-w_l|^2)/2}\\
&\times\frac{1}{(2\pi)^d} 
\frac{\hat{R}(\xi-P_0-\ldots-P_{N_{cr}-1})}
{\g(\xi-\ldots-P_{N_{cr}-1})-i\mathfrak{s}_{N_{cr}}|\xi-\ldots-P_{N_{cr}-1}|^2/2}
\prod_{l\in \mathcal{A}_{N_{cr}}\cup \mathcal{B}_{N_{cr}}}
\frac{1}{(2\pi)^d}\frac{\hat{R}(w_l)}{\g(w_l)+i\tau_l|w_l|^2/2}\\
&\times |\hat{\phi}_0(P_{N_{cr}})|^2\prod_{j=0}^{N_{cr}} 
\frac{(v_j-v_{j+1})^{|\mathcal{A}_j|+|\mathcal{B}_j|}}{(|\mathcal{A}_j|+|\mathcal{B}_j|)!}.
\end{aligned}
\end{equation*}
Here, we have changed the notation $r_i^-\mapsto v_i$, with
$v_0=t,v_{N_{cr}+1}=0$. Next, we integrate out~$w_l$ except for
$P_1,\ldots,P_{N_{cr}}$, so that
\begin{equation}
\begin{aligned}
&J_{m,n^*}^\eps(\sigma,\F_\sigma,\xi,F)\\
\to&(-1)^{n-k}\int_{\Delta_{N_{cr}}(t)} dv\int_{\R^{N_{cr}d}} dP
\prod_{j=1}^{N_{cr}-1}\frac{1}{(2\pi)^d
}\frac{\hat{R}(P_j)}{g(P_j)-i\mathfrak{s}_j(|\xi-\ldots-P_{j-1}|^2-|\xi-\ldots-P_j|^2)/2}\\
&\times  \prod_{j=0}^{N_{cr}-1}
(D(\xi-\ldots-P_j)/2)^{|\mathcal{A}_j|}
(D^*(\xi-\ldots-P_j)/2)^{|\mathcal{B}_j|}(D(0)/2)^{|\mathcal{A}_{N_{cr}}|}
(D^*(0)/2)^{|\mathcal{B}_{N_{cr}}|}\\
&\times \frac{\hat{R}(\xi-\ldots-P_{N_{cr}-1})}
{g(\xi-\ldots-P_{N_{cr}-1})-i\mathfrak{s}_{N_{cr}}|\xi-\ldots-P_{N_{cr}-1}|^2/2} 
|\hat{\phi}_0(P_{N_{cr}})|^2\prod_{j=0}^{N_{cr}}
\frac{(v_j-v_{j+1})^{|\mathcal{A}_j|+|\mathcal{B}_j|}}
{(|\mathcal{A}_j|+|\mathcal{B}_j|)!}
\label{eq:2mTrc3}
\end{aligned}
\end{equation}
Therefore, we have 
\[
\lim_{\eps\to 0}\E\{|\Psi_\eps(t,\xi)|^2\}=
\sum_{m,n=0}^\infty \sum_{\sigma:N_{cr}\geq 1}\lim_{\eps\to 0}J_{m,n^*}^\eps(\sigma,\F_\sigma,\xi,F)
\] 
with 
\[
\lim_{\eps\to 0}J_{m,n^*}^\eps(\sigma,\F_\sigma,\xi,F)
\]
given by the RHS of \eqref{eq:2mTrc3}.  It is clear that $n-N_{cr}$ is
even, so that $(-1)^{n-k}=(-1)^{k-N_{cr}}$ and we also note that
\[
k-N_{cr}=\sum_{i=0}^{N_{cr}}(|\mathcal{A}_i|+|\mathcal{B}_i|).
\]
When those crossing edges and $|\mathcal{A}_j|,|\mathcal{B}_j|$ are
fixed for $j=0,\ldots,N_{cr}$ (so the RHS of \eqref{eq:2mTrc3} is
fixed), the total number of possible permutations is
\[
\prod_{j=0}^{N_{cr}}\frac{(|\mathcal{A}_j|+|\mathcal{B}_j|)!}{|\mathcal{A}_j|!|\mathcal{B}_j|!}.
\]
Now, we can sum over all permutations when $N_{cr}$ is fixed, denoted
by $\sigma_{N_{cr}}$, and integrate in $P_{N_{cr}}$ and obtain
\begin{equation}
\begin{aligned}
&\sum_{\sigma_{N_{cr}}}\lim_{\eps\to 0}J_{m,n^*}^\eps(\sigma,\F_\sigma,t,\xi)\\
=&\|\hat{\phi}_0\|_2^2\sum_{|\mathcal{A}_1|,|\mathcal{B}_1|=0}^\infty
\ldots\sum_{|\mathcal{A}_{N_{cr}}|,|\mathcal{B}_{N_{cr}}|=0}^\infty
\int_{\Delta_{N_{cr}}(t)} dv\int_{\R^{(N_{cr}-1)d}} dP 
\prod_{j=0}^{N_{cr}} \frac{(-(v_j-v_{j+1}))^{|\mathcal{A}_j|+|\mathcal{B}_j|}}{|\mathcal{A}_j|!|\mathcal{B}_j|!}\\
&\times \left( \prod_{j=1}^{N_{cr}-1}
  \red(P_j,\xi-\ldots-P_{j-1})\right)
\red(\xi-\ldots-P_{N_{cr}-1},\xi-\ldots-P_{N_{cr}-1})\\
&\times  \left(\prod_{j=0}^{N_{cr}-1}
  (D(\xi-\ldots-P_j)/2)^{|\mathcal{A}_j|}
(D^*(\xi-\ldots-P_j)/2)^{|\mathcal{B}_j|}\right)(D(0)/2)^{|\mathcal{A}_{N_{cr}}|}(D^*(0)/2)^{|\mathcal{B}_{N_{cr}}|}.
\end{aligned}
\end{equation}
After the summation, we get
\begin{equation}
\begin{aligned}
&\sum_{\sigma_{N_{cr}}} \lim_{\eps\to 0}J_{m,n^*}^\eps(\sigma,\F_\sigma,t,\xi)
=\|\hat{\phi}_0\|_2^2\int_{\Delta_{N_{cr}}(t)}
dv\int_{\R^{(N_{cr}-1)d}} dP
\left(\prod_{j=0}^{N_{cr}-1}e^{-(v_j-v_{j+1})\red(\xi-\ldots-P_j)}\right)
\\
&\times e^{-v_{N_{cr}}\red(0)}
\left( \prod_{j=1}^{N_{cr}-1} \red(P_j,\xi-\ldots-P_{j-1})\right)\red(\xi-\ldots-P_{N_{cr}-1},\xi-\ldots-P_{N_{cr}-1}),
\end{aligned}
\end{equation}
which can also be written as
\begin{equation}
\begin{aligned}
\sum_{\sigma_{N_{cr}}} \lim_{\eps\to 0}J_{m,n^*}^\eps(\sigma,\F_\sigma,t,\xi)=&\|\hat{\phi}_0\|_2^2\int_{\Delta_{N_{cr}}(t)} dv\int_{\R^{N_{cr}d}} dP \left(\prod_{j=0}^{N_{cr}}e^{-(v_j-v_{j+1})\red(\xi-\ldots-P_j)}\right) \\
&\times \left( \prod_{j=1}^{N_{cr}} \red(P_j,\xi-\ldots-P_{j-1})\right)\delta(\xi-P_1-\ldots-P_{N_{cr}}).
\end{aligned}
\end{equation}
Thus, we have
\begin{equation}
\begin{aligned}
\lim_{\eps\to 0}\E\{|\Psi_\eps(t,\xi)|^2\}=&\|\hat{\phi}_0\|_2^2\sum_{N_{cr}=1}^\infty\int_{\Delta_{N_{cr}}(t)} dv\int_{\R^{N_{cr}d}} dP \left(\prod_{j=0}^{N_{cr}}e^{-(v_j-v_{j+1})\red(\xi-\ldots-P_j)}\right)\\
&\times \left( \prod_{j=1}^{N_{cr}} \red(P_j,\xi-\ldots-P_{j-1})\right)\delta(\xi-P_1-\ldots-P_{N_{cr}})=\widehat{W}_\delta(t,\xi).
\end{aligned}
\end{equation}
The proof of Lemma~\ref{lem:con2mTr} is complete.


\subsubsection*{Convergence of the higher order moments}

In this section, we consider convergence of the general moments
\[
\E\{\Psi_\eps(t,\xi)^M(\Psi_\eps^*(t,\xi))^N\},
\]
for arbitrary
$M,N\in\mathbb{N}$. By Lemma~\ref{lem:gnbd}, we can write
\begin{equation}
\E\{\Psi_\eps(t,\xi)^M(\Psi_\eps^*(t,\xi))^N\}=\sum_{m_1,\ldots,n_N=0}^\infty \E\{g_{m_1,\eps}\ldots g_{m_M,\eps} g_{n_1,\eps}^*\ldots g_{n_N,\eps}^*\}
\end{equation}
with $g_{n,\eps}(t,\xi)=F_{n,\eps}(t,\xi)$. As for the variance, we
only need to consider 
\[
\lim_{\eps\to   0}J_{m_1,\ldots,n_N^*}^\eps(\sigma,\F_\sigma,\xi,F),
\]
for fixed
$m_1,\ldots,n_N$ and $\sigma$ such that $N_c(\F_\sigma)\leq 2$.
Recall that \eqref{eq:defJ} gives
\begin{equation*}
\begin{aligned}
&|J_{m_1,\ldots,n_N^*}^\eps(\sigma,\F_\sigma,\xi,F)|\\
\leq &\frac{1}{\eps^{2k}}\int_{\sigma_{2k}(t)} dsdu \int_{\R^{2kd}} 
\prod_{(v_l,v_r)\in\F_\sigma}(2\pi)^{-d} e^{-\g(w_l)|v_{l}-v_{r}|/\eps^2}
\delta(w_l+ w_r)\hat{R}(w_l)dw_ldw_r 
\prod_{i=1}^M  |h_{M,i}|\prod_{j=1}^N |h_{N,j}^*|.
\end{aligned}
\end{equation*}
As before, we denote
\[
A_i=\{s_{i,1},\ldots,s_{i,m_i}\},~~~
B_j=\{u_{j,1},\ldots,u_{j,n_j}\},
\hbox{ with $i=1,\ldots,M$, $j=1,\ldots,N$.}
\]
The pairing $\F_\sigma$ decomposes 
\[
\{A_i,B_j:i=1,\ldots,M,j=1,\ldots,N\}
\]
into the connected components. If there exists a component of size
one, that is, $N_s(\sigma)=1$, then, as in the proof of
Lemma~\ref{lem:con1mTr}, we have a factor of
\[
\eps^{-\alpha
  d/2}|\hat{\phi}_0|(\xi/\eps^\alpha),
\]
coming from the corresponding initial condition, which implies
that
\[
J_{m_1,\ldots,n_N^*}^\eps(\sigma,\F_\sigma,\xi,F)\to 0
\] 
as $\eps\to 0$ since $\xi\neq 0$. 

Thus, we only need to consider the case when
\[
N_s(\sigma)=N_c(\sigma)=2.
\]
For any $S_1,S_2\in
\{A_i,B_j:i=1,\ldots,M,j=1,\ldots,N\}$ such that $S_1\leftrightarrow
S_2$, the following lemma shows that $S_1,S_2$ can not be of the same type.
\begin{lemma}
Fix $\sigma$ and assume $N_c(\sigma)=2$. 
If there exists a pair $S_1,S_2\in\{ A_i:i=1,\ldots,M\}$ 
or~$S_1,S_2\in\{B_j:j=1,\ldots,N\}$ such that $S_1\leftrightarrow S_2$, 
then 
\[
\lim_{\eps\to 0}J_{m_1,\ldots,n_N^*}^\eps(\sigma,\F_\sigma,\xi,F)=0.
\]
\label{lem:diffTr}
\end{lemma}
{\bf Proof.}
Let us assume that $S_1=A_{i_1},S_2=A_{i_2}$ -- the proof for the other case is
identical. Then we can write
\begin{equation*}
\begin{aligned}
|J_{m_1,\ldots,n_N^*}^\eps(\sigma,\F_\sigma,\xi,F)|
\leq &\frac{1}{\eps^{2k}}\int_{\sigma_{2k}(t)} dsdu \int_{\R^{2kd}} 
\prod_{(v_l,v_r)\in\F_\sigma}  
e^{-\g(w_l)|v_{l}-v_{r}|/\eps^2}\delta(w_l+ w_r)\farc{\hat{R}(w_l)}{(2\pi)^d}dw_ldw_r \\
& \times
|h_{M,i_1}h_{M,i_2}|\prod_{i=1, i\neq i_1,i_2}^M  |h_{M,i}|\prod_{j=1}^N |h_{N,j}^*|.
\end{aligned}
\end{equation*}
Since $A_{i_1}\leftrightarrow A_{i_2}$ and $N_{cr}(\sigma)=2$, after
integrating in $w_r$, we have 
\[
h_{M,i_1}=\eps^{-\alpha
  d/2}\hat{\phi}_0(\frac{\xi-P}{\eps^\alpha}),\hbox{ and }
h_{M,i_2}=\eps^{-\alpha d/2}\hat{\phi}_0(\frac{\xi+P}{\eps^\alpha}),
\]
for some variable 
\[
P=\sum_j p_{i_1,j}\neq 0,
\]
where the range of $j$ in
the summation depends on $\sigma$. Now we only need to pick some
$p_{i_1,j}$ and change this variable so that
$({\xi-P}){\eps^\alpha}\mapsto P$, which leads to
\begin{equation}
|h_{M,i_1}h_{M,i_2}|dp_{i_1,j}\mapsto \eps^{-\alpha d/2}
|\hat{\phi}_0(P)|\eps^{-\alpha
  d/2}|\hat{\phi}_0(\frac{2\xi}{\eps^\alpha}-P)|
\eps^{\alpha d}dP=|\hat{\phi}_0(P)\hat{\phi}_0(\frac{2\xi}{\eps^\alpha}-P)|dP.
\end{equation}
Then we perform the change of variables as in the proof of
Lemma~\ref{lem:bdmm} 
for 
\[
\prod_{i=1, i\neq i_1,i_2}^M  |h_{M,i}|\prod_{j=1}^N |h_{N,j}^*|,
\]
and in the end obtain
\begin{equation*}
\begin{aligned}
&|J_{m_1,\ldots,n_N^*}^\eps(\sigma,\F_\sigma,\xi,F)|\\
\leq & \frac{C^{M+N}}{\eps^{2k}}\int_{[0,t]^{2k}} dsdu 
\int_{\R^{kd}} \prod_{(v_l,v_r)\in\F_{\sigma,1}}  
e^{-\g(w_l)|v_{l}-v_{r}|/\eps^2}\frac{\hat{R}(w_l)}{(2\pi)^{d}}
\prod_{(v_l,v_r)\in\F_{\sigma,2}}(2\pi)^{-d} e^{-\g(z_l)|v_{l}-v_{r}|/\eps^2}\hat{R}(z_l)\\
 &\times
|\hat{\phi}_0(\tilde{w})\hat{\phi}_0(\frac{2\xi}{\eps^\alpha}-\tilde{w})|
\prod_{(v_l,v_r)\in\tilde{\F}_{\sigma,2}}|\hat{\phi}_0(w_l)|dwd\tilde{w}.
\end{aligned}
\end{equation*}
Here, as previously, $z_l$ denotes some momentum variables -- we will
not need their precise form, while~$(v_l,v_r)\in\F_{\sigma,1}$ denotes
the pairings not affected by the change of variables,
and~$(v_l,v_r)\in\F_{\sigma,2}$ denotes the affected
pairings. Finally, $\tilde{\F}_{\sigma,2}$ corresponds to the affected
pairings when we change variables for 
\[
\prod_{i=1, i\neq i_1,i_2}^M
|h_{M,i}|\prod_{j=1}^N |h_{N,j}^*|,
\]
as in the proof of aforementioned Lemma~\ref{lem:bdmm}.  We have also
changed the notation $P\mapsto \tilde{w}$. Now, after the temporal
integration we can apply dominated convergence theorem to obtain
\[
J_{m_1,\ldots,n_N^*}^\eps(\sigma,\F_\sigma,\xi,F)\to 0,
\]
due to the factor
$\hat{\phi}_0(\frac{2\xi}{\eps^\alpha}-\tilde{w})$.~$\Box$

By the above discussion, the nontrivial contribution of
$J_{m_1,\ldots,n_N^*}^\eps(\sigma,\F_\sigma,\xi,F)$ as $\eps \to 0$
comes only from the cases when $M=N$, the permutation $\sigma$ is such
that 
\[
N_s(\sigma)=N_c(\sigma)=2,
\]
and all connected components contain both type-$A$ and type-$B$
sets. Let $\Sigma(m_1,\ldots,n_N^*)$ be the set of such permutations.
For $\sigma\in\Sigma(m_1,\ldots,n_N^*)$, we have $A_i\leftrightarrow
B_{\tilde{i}}, i=1,\ldots,M$, where $\{\tilde{1},\ldots,\tilde{M}\}$
is a permutation of $\{1,\ldots,M\}$. We denote the set of $\sigma$
corresponding to a given $\{\tilde{1},\ldots,\tilde{M}\}$
by~$\Sigma_{\{\tilde{1},\ldots,\tilde{M}\}}(m_1,\ldots,n_N^*)$.
It is straightforward to check that
\begin{equation*}
\begin{aligned}
\sum_{\sigma}
1_{\sigma\in\Sigma_{\{\tilde{1},\ldots,\tilde{M}\}}(m_1,\ldots,n_N^*)}
J_{m_1,\ldots,n_N^*}^\eps(\sigma,\F_\sigma,\xi,F)
=\sum_{\sigma_{m_1,n_{\tilde{1}}^*}}\ldots\sum_{\sigma_{m_M,n_{\tilde{M}}^*}}
\prod_{i=1}^M
J_{m_i,n_{\tilde{i}}^*}^\eps(\sigma_{m_i,n_{\tilde{i}}^*}, 
\F_{\sigma_{m_i,n_{\tilde{i}}^*}},\xi,F),
\end{aligned}
\end{equation*}
where $\sigma_{m_i,n_{\tilde{i}}^*}$ denotes the permutation of
$A_i\cup B_{\tilde{i}}$ which keeps $A_i\leftrightarrow
B_{\tilde{i}}$. 
Now, we can write
\begin{equation*}
\begin{aligned}
&\lim_{\eps\to 0}\E\{\Psi_\eps(t,\xi)^M(\Psi_\eps^*(t,\xi))^N\}\\
=&\sum_{m_1,\ldots,n_N=0}^\infty  \sum_{\{\tilde{1},\ldots,\tilde{M}\}} \sum_{\sigma}1_{\sigma\in\Sigma_{\{\tilde{1},\ldots,\tilde{M}\}}(m_1,\ldots,n_N^*)} \lim_{\eps\to 0}J_{m_1,\ldots,n_N^*}^\eps(\sigma,\F_\sigma,\xi,F)\\
=&\sum_{m_1,\ldots,n_N=0}^\infty \sum_{\{\tilde{1},\ldots,\tilde{M}\}} \sum_{\sigma_{m_1,n_{\tilde{1}}^*}}\ldots\sum_{\sigma_{m_M,n_{\tilde{M}}^*}}\prod_{i=1}^M \lim_{\eps\to 0} J_{m_i,n_{\tilde{i}}^*}^\eps(\sigma_{m_i,n_{\tilde{i}}^*}, \F_{\sigma_{m_i,n_{\tilde{i}}^*}},\xi,F)\\
=& \sum_{\{\tilde{1},\ldots,\tilde{M}\}} \prod_{i=1}^M \left( \sum_{m_i,n_{\tilde{i}}^*=0}^\infty\sum_{\sigma_{m_i,n_{\tilde{i}}^*}} \lim_{\eps\to0}  J_{m_i,n_{\tilde{i}}^*}^\eps(\sigma_{m_i,n_{\tilde{i}}^*},\F_{\sigma_{m_i,n_{\tilde{i}}^*}},\xi,F)\right)
= M! \widehat{W}_\delta(t,\xi)^M.
\end{aligned}
\end{equation*}
Here, the last equality comes from Lemma~\ref{lem:con2mTr}: 
\[
\lim_{\eps\to 0}\E\{|\Psi_\eps(t,\xi)|^2\}=\sum_{m_i,n_{\tilde{i}}^*=0}^\infty \sum_{\sigma_{m_i,n_{\tilde{i}}^*}} \lim_{\eps\to0}  J_{m_i,n_{\tilde{i}}^*}^\eps(\sigma_{m_i,n_{\tilde{i}}^*},\F_{\sigma_{m_i,n_{\tilde{i}}^*}},\xi,F)=\widehat{W}_\delta(t,\xi).
\]
To summarize, we have shown that 
\[
\lim_{\eps\to
  0}\E\{\Psi_\eps(t,\xi)^M(\Psi_\eps^*(t,\xi))^N\}=1_{M=N}M!
\widehat{W}_\delta(t,\xi)^M,
\]
for arbitrary $M,N\in\mathbb{N}$. The
proof of Theorem~\ref{thm:trHigh} is complete.

\section{The fluctuation analysis}\label{sec:fluct}

In this section, we prove Theorems~\ref{thm:cor} and~\ref{thm:wig}.

\subsection*{Pointwise fluctuation}

We begin with Theorem~\ref{thm:cor}.
Recall that the corrector can be written as
\[
\U_\eps(t,\xi)=\eps^{-\alpha d/2}\sum_{n=0}^\infty
\FF_{n,\eps}(t,\xi),
\]
and we have previously shown that 
\begin{equation}
\begin{aligned}
\lim_{\eps\to 0}\eps^{-\alpha
  d(M+N)/2}\E\{\FF_{m_1,\eps}\ldots\FF_{m_M,\eps}
\FF_{n_1,\eps}^*\ldots\FF_{n_N,\eps}^*\}
=\!\!\!
\sum_{\sigma:N_s(\F_\sigma)=N_c(\F_\sigma)=2} \!
\lim_{\eps\to 0}J_{m_1,\ldots,n_N^*}^\eps(\sigma,\F_\sigma,\eps^\alpha \xi,F),
\end{aligned}
\end{equation}
when $M+N=2K$ for some $K\in\mathbb{N}$.
Let us define 
\[
\tilde{\Sigma}(m_1,\ldots,n_N^*)=\{\sigma:N_s(\F_\sigma)=N_c(\F_\sigma)=2\}.
\]
The constraint 
\[
N_s(\F_\sigma)=N_c(\F_\sigma)=2
\]
forms pairings over vertices
\begin{equation}
\{C_l:l=1,\ldots,M+N\}=\{A_i,B_j:i=1,\ldots,M,j=1,\ldots,N\},
\end{equation}
or equivalently the set
\begin{equation}\label{may302}
\{m_i,n_j^*: i=1,\ldots,M,j=1,\ldots,N\}.
\end{equation}
We write 
\[
\tilde{\Sigma}(m_1,\ldots,n_N^*)=
\bigcup_{\mathfrak{p}} \tilde{\Sigma}_\mathfrak{p}(m_1,\ldots,n_N^*),
\]
where $\tilde{\Sigma}_\mathfrak{p}(m_1,\ldots,n_N^*)$ is the
set of permutations corresponding to a given pairing $\mathfrak{p}$
over~(\ref{may302}). 
Then we can write
\begin{equation*}
\sum_{\sigma:N_s(\F_\sigma)=N_c(\F_\sigma)=2} \lim_{\eps\to 0}J_{m_1,\ldots,n_N^*}^\eps(\sigma,\F_\sigma,\eps^\alpha \xi,F)=\sum_{\mathfrak{p}} \sum_{\sigma\in\tilde{\Sigma}_{\mathfrak{p}}(m_1,\ldots,n_N^*)} \lim_{\eps\to 0}J_{m_1,\ldots,n_N^*}^\eps(\sigma,\F_\sigma,\eps^\alpha \xi,F).
\end{equation*}
For a given $\mathfrak{p}$, we assume that pairs have the 
form $(\PF(l),\PF(\tilde{l}))$ with $l=1,\ldots,K$, where 
\begin{equation}
\{\PF(l),\PF(\tilde{l}):l=1,\ldots,K\}=\{m_i,n_j^*: i=1,\ldots,M,j=1,\ldots,N\}.
\end{equation}
It is straightforward to check that
\begin{equation}
\begin{aligned}
&\sum_{\sigma\in\tilde{\Sigma}_{\mathfrak{p}}(m_1,\ldots,n_N^*)} 
J_{m_1,\ldots,n_N^*}^\eps(\sigma,\F_\sigma,\eps^\alpha \xi,F)\\
=&\sum_{\sigma(\PF(1),\PF(\tilde{1}))}\ldots\sum_{\sigma(\PF(K),\PF(\tilde{K}))} 
\prod_{l=1}^{K}J_{\PF(l),\PF(\tilde{l})}^\eps(\sigma(\PF(l),\PF(\tilde{l})),
\F_{\sigma(\PF(l),\PF(\tilde{l}))},\eps^\alpha \xi,F),
\end{aligned}
\end{equation}
where $\sigma(\PF(l),\PF(\tilde{l}))$ denotes the permutation of
$C_i\cup C_j$ such that $C_i\leftrightarrow C_j$ if
$\PF(l),\PF(\tilde{l})$ corresponds to $C_i,C_j$.
Now, we have
\begin{equation}
\begin{aligned}
&\lim_{\eps\to 0}\E\{\U_\eps(t,\xi)^M(\U_\eps^*(t,\xi))^N\}
=\sum_{m_1,\ldots,n_N=0}^\infty \sum_{\mathfrak{p}} 
\sum_{\sigma\in\tilde{\Sigma}_{\mathfrak{p}}(m_1,\ldots,n_N^*)} 
\lim_{\eps\to 0}J_{m_1,\ldots,n_N^*}^\eps(\sigma,\F_\sigma,t,\eps^\alpha \xi)\\
=&\sum_{m_1,\ldots,n_N=0}^\infty
\sum_{\mathfrak{p}}\sum_{\sigma(\PF(1),\PF(\tilde{1}))}
\ldots\sum_{\sigma(\PF(K),\PF(\tilde{K}))} 
\prod_{l=1}^{K}\lim_{\eps\to 0}
J_{\PF(l),\PF(\tilde{l})}^\eps(\sigma(\PF(l),\PF(\tilde{l})),
\F_{\sigma(\PF(l),\PF(\tilde{l}))},\eps^\alpha \xi,F)\\
=&\sum_{\mathfrak{p}} \prod_{l=1}^K
\left(\sum_{\PF(l),\PF(\tilde{l})=0}^\infty 
\sum_{\sigma(\PF(l),\PF(\tilde{l}))} 
\lim_{\eps\to 0}
J_{\PF(l),\PF(\tilde{l})}^\eps(\sigma(\PF(l),\PF(\tilde{l})),
\F_{\sigma(\PF(l),\PF(\tilde{l}))},\eps^\alpha \xi,F)\right).
\label{eq:corhighto2}
\end{aligned}
\end{equation}
Therefore, it is clear that we only need to compute
\begin{equation*}
\sum_{m,n=0}^\infty\sum_{\sigma} 
\lim_{\eps\to 0}J_{m,n^*}^\eps(\sigma,\F_\sigma,\eps^\alpha \xi,F) 
\mbox{ and }\sum_{m,n=0}^\infty\sum_{\sigma} 
\lim_{\eps\to 0}J_{m,n}^\eps(\sigma,\F_\sigma,\eps^\alpha \xi,F)
\end{equation*}
to obtain $\lim_{\eps\to
  0}\E\{\U_\eps(t,\xi)^M(\U_\eps^*(t,\xi))^N\}$. The following lemmas
combine to conclude the proof of Theorem~\ref{thm:cor}.

The first lemma deals with the ``complex-conjugate'' moments.
\begin{lemma}
We have
\[
\sum_{m,n=0}^\infty\sum_{\sigma} \lim_{\eps\to
  0}J_{m,n^*}^\eps(\sigma,\F_\sigma,\eps^\alpha
\xi,F)=\widehat{W}_{\delta,s}(t,0).
\]
\label{lem:corab2m}
\end{lemma}
{\bf Proof.}
Following the proof of Lemma~\ref{lem:con2mTr} with $\xi$ replaced by
$\eps^\alpha \xi$, we obtain
\begin{equation*}
\begin{aligned}
&\sum_{m,n=0}^\infty\sum_{\sigma} 
\lim_{\eps\to 0}J_{m,n^*}^\eps(\sigma,\F_\sigma,t,\eps^\alpha \xi)\to
\|\hat{\phi}_0\|_2\sum_{N_{cr}=1}^\infty \int_{\Delta_{N_{cr}}(t)}
dv\int_{\R^{N_{cr}d}} dP \\
&\times\left(\prod_{j=0}^{N_{cr}}e^{-(v_j-v_{j+1})\red(-P_0-\ldots-P_j)}\right)
 \left( \prod_{j=1}^{N_{cr}} \red(P_j,-P_0-\ldots-P_{j-1})\right)\delta(-P_1-\ldots-P_{N_{cr}}).
\end{aligned}
\end{equation*}
The RHS equals to $\widehat{W}_{\delta,s}(t,0)$, which completes the proof.~$\Box$

The second lemma address the ``non-conjugated'' moments.
\begin{lemma}
\begin{equation*}
\begin{aligned}
\sum_{m,n=0}^\infty\sum_{\sigma} \lim_{\eps\to 0}J_{m,n}^\eps(\sigma,\F_\sigma,\eps^\alpha \xi,F)=\mathcal{W}_\alpha(t,\xi).
\end{aligned}
\end{equation*}
\label{lem:cor2m}
\end{lemma} 

{\bf Proof}. We use the same notation in the proof of
Lemma~\ref{lem:con2mTr}.
Recall that
\[
\begin{aligned}
&J_{m,n}^\eps(\sigma,\F_\sigma,\eps^\alpha \xi,F)\\
=&\frac{1}{(i\eps)^{m+n}}\int_{\sigma_{2k}(t)} dsdu\int_{\R^{2kd}} 
\prod_{(v_l,v_r)\in\F_\sigma}(2\pi)^{-d}
e^{-\g(w_l)|v_{l}-v_{r}|/\eps^2}\delta(w_l+ w_r)
\hat{R}(w_l)dw_ldw_r\\
&\times e^{iG_m( \eps^\alpha \xi, s^{(m)},p^{(m)})/\eps^{2}}
e^{iG_n(\eps^\alpha \xi, u^{(n)},q^{(n)})/\eps^{2}} \eps^{-\alpha d}
\hat{\phi}_0(\frac{\eps^\alpha \xi-p_1-\ldots-p_m}{\eps^\alpha})
\hat{\phi}_0(\frac{\eps^\alpha \xi-q_1-\ldots-q_n}{\eps^\alpha}).
\end{aligned}
\]
We only need to consider $\sigma$ such that the number of crossing edges
$N_{cr}\geq 1$. For each crossing
edge~$(r_i^+,r_i^-),i=1,\ldots,N_{cr}$, we denote the $p-$variable
by $P_i$. After the integration of the
delta functions, we obtain
\begin{equation}
\begin{aligned}
&J_{m,n}^\eps(\sigma,\F_\sigma,\eps^\alpha \xi,F)
=\frac{1}{(i\eps)^{m+n}}\int_{\sigma_{2k}(t)} dsdu\int_{\R^{kd}} dw\\
&\times\prod_{j=1}^{k}\frac{\hat{R}(w_j)}{(2\pi)^d}e^{-\g(w_j)\frac{v_j^+-v_j^-}{\eps^2}}\prod_{j=1}^{N_{cr}} e^{i(|\eps^\alpha \xi-\mathfrak{s}_j(P_0+\ldots+P_{j-1})|^2-|\eps^\alpha \xi-\mathfrak{s}_j(P_0+\ldots+P_j)|^2)\frac{r_j^+-r_j^-}{2\eps^2}}\\
&\times \prod_{j=0}^{N_{cr}}\left(\prod_{l\in \mathcal{A}_j}e^{i(|\eps^\alpha \xi-\ldots-P_j|^2-|\eps^\alpha \xi-\ldots-P_j-w_l|^2)\frac{v_l^+-v_l^-}{2\eps^2}} \prod_{l\in  \mathcal{B}_j}e^{i(|\eps^\alpha \xi+\ldots+P_j|^2-|\eps^\alpha \xi+\ldots+P_j-w_l|^2)\frac{v_l^+-v_l^-}{2\eps^2}}\right)\\
&\times  \prod_{j=1}^{N_{cr}} e^{-i(|P_j|^2+2P_j\cdot (P_0+\ldots+P_{j-1}))\frac{r_j^-}{\eps^2}}\frac{1}{\eps^{\alpha d}}\hat{\phi}_0(\frac{\eps^\alpha \xi-P_0-\ldots-P_{N_{cr}}}{\eps^\alpha})\hat{\phi}_0(\frac{\eps^\alpha \xi+P_0+\ldots+P_{N_{cr}}}{\eps^\alpha}). 
\label{eq:coreq1}
\end{aligned}
\end{equation}
Compared to \eqref{eq:2mTrc2}, the key difference is that we get an
extra factor with a large phase: 
\[
\prod_{j=1}^{N_{cr}}
e^{-i(|P_i|^2+2P_i\cdot
  (P_0+\ldots+P_{i-1}))r_j^-/\eps^2}.
\]
To get rid of the factor ${\eps^{-\alpha d}}$, we change the
variable 
\[
P_{N_{cr}}\mapsto -P_0-\ldots-P_{N_{cr}-1}+\eps^\alpha
P_{N_{cr}}.
\]
Rewriting the terms in \eqref{eq:coreq1} associated with
$P_{N_{cr}}$ using the new variable gives
\begin{equation}
\begin{aligned}
&J_{m,n}^\eps(\sigma,\F_\sigma,\eps^\alpha \xi,F)=\frac{1}{(i\eps)^{2k}}\int_{\sigma_{2k}(t)} dsdu\int_{\R^{kd}} dw\\
&\prod_{j:w_j\neq P_{N_{cr}}}\frac{\hat{R}(w_j)}{(2\pi)^d}e^{-\g(w_j)\frac{v_j^+-v_j^-}{\eps^2}}
\frac{\hat{R}(-P_0-\ldots-P_{N_{cr}-1}+\eps^\alpha P_{N_{cr}})}{(2\pi)^d}e^{-\g(-P_0-\ldots-P_{N_{cr}-1}+\eps^\alpha P_{N_{cr}})\frac{r_{N_{cr}}^+-r_{N_{cr}}^-}{\eps^2}}\\
&\times \prod_{j=1}^{N_{cr}-1} \left(e^{i(|\eps^\alpha
    \xi-\mathfrak{s}_j(P_0+\ldots+P_{j-1})|^2-|\eps^\alpha
    \xi-\mathfrak{s}_j(P_0+\ldots+P_j)|^2)\frac{r_j^+-r_j^-}{2\eps^2}}\right)\\
&\times 
e^{i(|\eps^\alpha \xi-\mathfrak{s}_{N_{cr}}(P_0+\ldots+P_{N_{cr}-1})|^2-|\eps^\alpha \xi-\mathfrak{s}_{N_{cr}}\eps^\alpha P_{N_{cr}}|^2)\frac{r_{N_{cr}}^+-r_{N_{cr}}^-}{2\eps^2}}\\
&\times \prod_{j=0}^{N_{cr}-1}\left(\prod_{l\in \mathcal{A}_j}e^{i(|\eps^\alpha \xi-\ldots-P_j|^2-|\eps^\alpha \xi-\ldots-P_j-w_l|^2)\frac{v_l^+-v_l^-}{2\eps^2}} \prod_{l\in  \mathcal{B}_j}e^{i(|\eps^\alpha \xi+\ldots+P_j|^2-|\eps^\alpha \xi+\ldots+P_j-w_l|^2)\frac{v_l^+-v_l^-}{2\eps^2}}\right)\\
&\times \prod_{l\in \mathcal{A}_{N_{cr}}}e^{i(|\eps^\alpha\xi-\eps^\alpha P_{N_{cr}}|^2-|\eps^\alpha\xi-\eps^\alpha P_{N_{cr}}-w_l|^2)\frac{v_l^+-v_l^-}{2\eps^2}} \prod_{l\in  \mathcal{B}_{N_{cr}}}e^{i(|\eps^\alpha \xi+\eps^\alpha P_{N_{cr}}|^2-|\eps^\alpha \xi+\eps^\alpha P_{N_{cr}}-w_l|^2)\frac{v_l^+-v_l^-}{2\eps^2}}\\
&\times  \prod_{j=1}^{N_{cr}-1} e^{-i(|P_j|^2+2P_j\cdot (P_0+\ldots+P_{j-1}))\frac{r_j^-}{\eps^2}}\hat{\phi}_0(\xi-P_{N_{cr}})\hat{\phi}_0(\xi+P_{N_{cr}})\\
&\times e^{-i(|-P_0-\ldots-P_{N_{cr}-1}+\eps^\alpha P_{N_{cr}}|^2+2(-P_0-\ldots-P_{N_{cr}-1}+\eps^\alpha P_{N_{cr}})\cdot (P_0+\ldots+P_{N_{cr}-1}))\frac{r_{N_{cr}}^-}{\eps^2}}. 
\label{eq:coreq2}
\end{aligned}
\end{equation}
If we freeze $r_1^-,\ldots,r_{N_{cr}}^-,P_1,\ldots,P_{N_{cr}}$,
integrate out the other variables, and send $\eps\to 0$, we see that
\[
\lim_{\eps\to 0} |J_{m,n}^\eps(\sigma,\F_\sigma,\eps^\alpha\xi,F)-
H_{m,n}^\eps(\sigma,\F_\sigma,\eps^\alpha\xi,F)|=0,
\]
with 
\begin{equation}
\begin{aligned}
& H_{m,n}^\eps(\sigma,\F_\sigma,\eps^\alpha\xi,F)=
\frac{1}{(-1)^k}\int_{\Delta_{N_{cr}}(t)} dv\int_{\R^{N_{c}d}} dP\\
 & \left(\prod_{j=1}^{N_{cr}-1}D(P_j,-P_0-\ldots-P_{j-1})/2\right)
D(-P_0-\ldots-P_{N_{cr}-1},-P_0-\ldots-P_{N_{cr}-1}))/2\\
 &\times \left( \prod_{j=0}^{N_{cr}-1}
(D(P_0+\ldots+P_j)/2)^{|\mathcal{A}_j|+|\mathcal{B}_j|}
\frac{(v_j-v_{j+1})^{|\mathcal{A}_j|+|\mathcal{B}_j|}}{(|\mathcal{A}_j|+|\mathcal{B}_j|)!}
\right) (D(0)/2)^{|\mathcal{A}_{N_{cr}}|+|\mathcal{B}_{N_{cr}}|}\\
&\times
\frac{(v_{N_{cr}}-v_{N_{cr}+1})^{|\mathcal{A}_{N_{cr}}|+|\mathcal{B}_{N_{cr}}|}}
{(|\mathcal{A}_{N_{cr}}|+|\mathcal{B}_{N_{cr}}|)!}
\prod_{j=1}^{N_{cr}-1} e^{-i(|P_j|^2+2P_j\cdot 
(P_0+\ldots+P_{j-1}))\frac{v_j}{\eps^2}}
\hat{\phi}_0(\xi-P_{N_{cr}})\hat{\phi}_0(\xi+P_{N_{cr}})\\
&\times e^{-i(|-P_0-\ldots-P_{N_{cr}-1}+\eps^\alpha P_{N_{cr}}|^2+
2(-P_0-\ldots-P_{N_{cr}-1}+\eps^\alpha P_{N_{cr}})\cdot 
(P_0+\ldots+P_{N_{cr}-1}))\frac{v_{N_{cr}}}{\eps^2}}.
\label{eq:coreq3}
\end{aligned}
\end{equation}
Here, we used the property 
\[
D(\xi)=D(-\xi)\hbox{ and }D(p,\xi)=D(-p,-\xi).
\]
We will consider separately the cases $N_{cr}\ge 2$ and $N_{cr}=1$.

\textbf{Multiple scattering $N_{cr}\geq 2$}. 
When $N_{cr}\ge 2$, 
we have at least one oscillatory phase in \eqref{eq:coreq3}, since
\begin{equation}
\begin{aligned}
&\prod_{j=1}^{N_{cr}-1} e^{-i(|P_j|^2+2P_j\cdot (P_0+\ldots+P_{j-1}))\frac{v_j}{\eps^2}}\\
&\times e^{-i(|-P_0-\ldots-P_{N_{cr}-1}+\eps^\alpha P_{N_{cr}}|^2+2(-P_0-\ldots-P_{N_{cr}-1}+\eps^\alpha P_{N_{cr}})\cdot (P_0+\ldots+P_{N_{cr}-1}))\frac{v_{N_{cr}}}{\eps^2}}= e^{-i|P_1|^2\frac{v_1}{\eps^2}} X
\end{aligned}
\end{equation}
with $|X|= 1$ and independent of $v_1$. For the integral
in $v$, we have
\begin{equation}
\begin{aligned}
&|\int_{\Delta_{N_{cr}}(t)}dv 
\prod_{j=0}^{N_{cr}}(v_j-v_{j+1})^{|\mathcal{A}_j|+|\mathcal{B}_j|}e^{-i|P_1|^2\frac{v_1}{\eps^2}} X|\\
\leq & C\int_{\Delta_{N_{cr}-1}(t)}\prod_{j=2}^{N_{cr}}dv_j |
\int_{v_2}^t(t-v_1)^{|\mathcal{A}_0|+|\mathcal{B}_0|}
(v_1-v_2)^{|\mathcal{A}_1|+|\mathcal{B}_1|} e^{-i|P_1|^2\frac{v_1}{\eps^2}}dv_1|
\end{aligned}
\end{equation}
for some $C$. Applying the Riemann-Lebesgue lemma gives
\begin{equation}
 |\int_{v_2}^t(t-v_1)^{|\mathcal{A}_0|+|\mathcal{B}_0|}
(v_1-v_2)^{|\mathcal{A}_1|+|\mathcal{B}_1|} e^{-i|P_1|^2\frac{v_1}{\eps^2}}dv_1|\to 0,
\end{equation}
provided that $P_1\neq 0$. Thus, by the dominated convergence theorem, we obtain 
\begin{equation}
\int_{\Delta_{N_{cr}}(t)}dv 
\prod_{j=0}^{N_{cr}}(v_j-v_{j+1})^{|\mathcal{A}_j|+|\mathcal{B}_j|}e^{-i|P_1|^2\frac{v_1}{\eps^2}}
X
\to 0,
\end{equation}
when $P_1\neq 0$, which implies
\[
H_{m,n}^\eps(\sigma,\F_\sigma,\eps^\alpha\xi,F)\to 0,\hbox{ as
  $\eps\to 0$},
\]
if $N_{cr}\geq 2$.

\textbf{Single scattering $N_{cr}=1$}. When $N_{cr}=1$,
\eqref{eq:coreq3} simplifies to
 \begin{equation}
 \begin{aligned}
& H_{m,n}^\eps(\sigma,\F_\sigma,\eps^\alpha\xi,F)=\frac{D(0,0)}{2(-1)^k}\int_0^t dv\int_{\R^{d}} dP  \hat{\phi}_0(\xi-P)\hat{\phi}_0(\xi+P)e^{-i|\eps^\alpha P|^2\frac{v}{\eps^2}}\\
 &\times  \left(\frac{D(0)}{2}\right)^{|\mathcal{A}_0|+|\mathcal{B}_0|+|\mathcal{A}_{1}|+|\mathcal{B}_{1}|}\frac{(t-v)^{|\mathcal{A}_0|+|\mathcal{B}_0|}}{(|\mathcal{A}_0|+|\mathcal{B}_0|)!}\frac{v^{|\mathcal{A}_{1}|+|\mathcal{B}_{1}|}}{(|\mathcal{A}_{1}|+|\mathcal{B}_{1}|)!}.
\end{aligned}
\end{equation}
If $\alpha\in(0,1)$, we have a large phase factor
$e^{i|P|^2v/\eps^{2-2\alpha}}$, so for the same reason as for
$N_{cr}\ge 2$, we have 
\[
H_{m,n}^\eps(\sigma,\F_\sigma,\eps^\alpha\xi,F)\to 0,
\] 
which implies 
\[
J_{m,n}^\eps(\sigma,\F_\sigma,\eps^\alpha\xi,F)\to 0.
\]
If $\alpha=1$, we have
 \begin{equation}
 \begin{aligned}
& H_{m,n}^\eps(\sigma,\F_\sigma,\eps^\alpha \xi,F)=\frac{D(0,0)}{2(-1)^k}\int_0^t dv\int_{\R^{d}} dP  \hat{\phi}_0(\xi-P)\hat{\phi}_0(\xi+P)e^{-i|P|^2v}\\
 &\times  \left(\frac{D(0)}{2}\right)^{|\mathcal{A}_0|+|\mathcal{B}_0|+|\mathcal{A}_{1}|+|\mathcal{B}_{1}|}\frac{(t-v)^{|\mathcal{A}_0|+|\mathcal{B}_0|}}{(|\mathcal{A}_0|+|\mathcal{B}_0|)!}\frac{v^{|\mathcal{A}_{1}|+|\mathcal{B}_{1}|}}{(|\mathcal{A}_{1}|+|\mathcal{B}_{1}|)!},
 \end{aligned}
 \end{equation}
which is $\eps-$independent. 
Following the argument we used in the proof of 
Lemma~\ref{lem:con2mTr}, we have
\begin{equation}
\begin{aligned}
\sum_\sigma \lim_{\eps\to
  0}J_{m,n}^\eps(\sigma,\F_\sigma,\eps^\alpha\xi,F)=
&\sum_{\sigma:N_{cr}=1}\lim_{\eps\to 0}H_{m,n}^\eps(\sigma,\F_\sigma,\eps^\alpha \xi,F)\\
=&-D(0,0)e^{-D(0)t}\int_0^tdv \int_{\R^d}dP\hat{\phi}_0(\xi-P)
\hat{\phi}_0(\xi+P)e^{-i|P|^2v}.
\label{eq:limH1}
\end{aligned}
\end{equation}
Finally, if $\alpha>1$, similarly, we have
\begin{equation}
\begin{aligned}
\sum_\sigma \lim_{\eps\to 0}J_{m,n}^\eps(\sigma,\F_\sigma,\eps^\alpha\xi,F)=&\sum_{\sigma:N_{cr}=1}\lim_{\eps\to 0}H_{m,n}^\eps(\sigma,\F_\sigma,\eps^\alpha\xi,F)\\
=&-D(0,0)e^{-D(0)t}t \int_{\R^d}dP\hat{\phi}_0(\xi-P)\hat{\phi}_0(\xi+P).
\label{eq:limH2}
\end{aligned}
\end{equation}
The proof of Lemma~\ref{lem:cor2m} is complete.

\begin{remark}
The proof shows that only single scattering contributes to the ``non-conjugated" moments when $\alpha\geq 1$. This is similar to the result obtained for heat equation \cite[Theorem 2]{B-MMS-10}, where the single scattering constitutes the whole random corrector. For Schr\"odinger equation, the situation is different, as multiple scatterings show up in ``complex-conjugated" moments as in the proof of Lemma~\ref{lem:corab2m}.
\end{remark}


\subsection*{Correlation of the fluctuations}

Here, we prove Theorem~\ref{thm:wig}. 
Recall that we look at the behavior of 
\begin{equation}
W_\eps(t,x,\xi)=\int_{\R^d}\U_\eps(t,\xi+\frac{\eps^\beta\eta}{2})
\U_\eps^*(t,\xi-\frac{\eps^\beta\eta}{2})e^{i\eta\cdot x} \frac{d\eta}{(2\pi)^d}.
\end{equation}
To prove the convergence of 
\[
\langle
W_\eps(t),\varphi\rangle=\int_{\R^d}W_\eps(t,x,\xi)\varphi^*(x,\xi)dxd\xi
\]
in probability, it suffices to show the convergence of
\[
\E\{\langle W_\eps(t),\varphi\rangle\}
\]
and
\[
\E\{|\langle W_\eps(t),\varphi\rangle|^2\}.
\]
Given that
\begin{equation*}
\begin{aligned}
\E\{\langle W_\eps(t),\varphi\rangle\}=&\frac{1}{(2\pi)^d}
\int_{\R^{3d}}\E\{\U_\eps(t,\xi+\frac{\eps^\beta\eta}{2})
\U_\eps^*(t,\xi-\frac{\eps^\beta\eta}{2})\}e^{i\eta\cdot
  x}\varphi^*(x,\xi)d\eta dxd\xi
\end{aligned}
\end{equation*}
and
\begin{equation*}
\begin{aligned}
\E\{|\langle W_\eps(t),\varphi\rangle|^2\}= 
\int_{\R^{6d}}&\E\{\U_\eps(t,\xi_1+\frac{\eps^\beta\eta_1}{2})
\U_\eps^*(t,\xi_1-\frac{\eps^\beta\eta_1}{2})
\U_\eps^*(t,\xi_2+\frac{\eps^\beta\eta_2}{2})
\U_\eps(t,\xi_2-\frac{\eps^\beta\eta_2}{2})\}\\
&\times e^{i\eta_1\cdot x_1}\varphi^*(x_1,\xi_1)
e^{-i\eta_2\cdot
  x_2}\varphi(x_2,\xi_2) \frac{d\eta dx d\xi}{(2\pi)^{2d}}  ,
\end{aligned}
\end{equation*}
we first prove the following two results. 
\begin{lemma}
If $\alpha+\beta=2$ and $\alpha\in(0,2]$, then as $\eps\to 0$, 
\begin{equation*}
\begin{aligned}
&\E\{\U_\eps(t,\xi+\frac{\eps^\beta\eta}{2})\U_\eps^*(t,\xi-\frac{\eps^\beta\eta}{2})\}\\
\to&\sum_{N_{cr}=1}^\infty\int_{\Delta_{N_{cr}}(t)}
dv\int_{\R^{N_{cr}d}} dP 
\left(\prod_{j=0}^{N_{cr}}e^{-(v_j-v_{j+1})\red(-P_0-\ldots-P_j)}\right) 
\left( \prod_{j=1}^{N_{cr}} \red(P_j,-P_0-\ldots-P_{j-1})\right)\\
&\times\delta(-P_1-\ldots-P_{N_{cr}})\prod_{j=1}^{N_{cr}}e^{iP_j\cdot
  \eta
  v_j}\left(1_{\alpha\in(0,2)}\|\hat{\phi}_0\|_2^2+1_{\alpha=2}
\int_{\R^d}\hat{\phi}_0(\xi+\frac{\eta}{2}-p)
\hat{\phi}_0^*(\xi-\frac{\eta}{2}-p)dp\right).
\end{aligned}
\end{equation*}
\label{lem:concrco}
\end{lemma}

\begin{lemma}
If $\xi_1\neq \xi_2$, $\alpha+\beta=2$ and $\alpha\in (0,1)$, then 
\begin{equation*}
\begin{aligned}
&\lim_{\eps\to 0}\E\{\U_\eps(t,\xi_1+\frac{\eps^\beta\eta_1}{2})
\U_\eps^*(t,\xi_1-\frac{\eps^\beta\eta_1}{2})
\U_\eps^*(t,\xi_2+\frac{\eps^\beta\eta_2}{2})\U_\eps(t,\xi_2-\frac{\eps^\beta\eta_2}{2})\}\\
=&\lim_{\eps\to 0}\E\{\U_\eps(t,\xi_1+\frac{\eps^\beta\eta_1}{2})
\U_\eps^*(t,\xi_1-\frac{\eps^\beta\eta_1}{2})\}
\E\{\U_\eps^*(t,\xi_2+\frac{\eps^\beta\eta_2}{2})\U_\eps(t,\xi_2-\frac{\eps^\beta\eta_2}{2})\}.
\end{aligned}
\end{equation*}
\label{lem:con4co}
\end{lemma}
The assumption $\alpha+\beta=2$ in Lemmas~\ref{lem:concrco} and
\ref{lem:con4co} matches the kinetic scaling. To see this, recall that
\[
\U_\eps(t,\xi)=\eps^{-\alpha  d/2}(\psi_\eps(t,\xi)-\E\{\psi_\eps(t,\xi)\}),
\]
and
\[
\psi_\eps(t,\xi)=\eps^{\alpha d}\hat{\phi}(t/\eps^2,\eps^\alpha \xi)
e^{i|\eps^\alpha \xi|^2t/2\eps^2}.
\]
If we let 
\[
\mathscr{U}(t,x)=\phi(t,x)-\E\{\phi(t,x)\},
\]
then the Wigner transform written in physical domain is
\[
\int_{\R^d}\mathscr{U}(\frac{t}{\eps^2},\frac{x}{\eps^{\alpha+\beta}}-\frac{y}{2\eps^\alpha})
\mathscr{U}^*(\frac{t}{\eps^2},\frac{x}{\eps^{\alpha+\beta}}+\frac{y}{2\eps^\alpha})e^{i\xi\cdot y}dy,
\]
that is, we need $\alpha+\beta=2$ so that the propagation speed is of
order one. Note the compensated phase factor from the compensation
\[
e^{i\xi\cdot \eta t \eps^{2\alpha+\beta-2}},
\]
disappears in the limit
when choosing $\alpha+\beta=2$.
 
{\bf Proof of Lemma~\ref{lem:concrco}.}
We will use the representation 
\[
\U_\eps(t,\xi)=\eps^{-\alpha d/2}\sum_{n\geq 1} \FF_{n,\eps}(t,\xi),
\]
so we only need to consider 
\[
\E\{\eps^{-\alpha
  d}\FF_{m,\eps}(t,\xi_{1})\FF_{n,\eps}^*(t,\xi_{-1})\},
\]
with 
\[
\xi_{1}=\xi+\frac{\eps^\beta\eta}{2}, ~~~\xi_{-1}=\xi-\frac{\eps^\beta
  \eta}{2}.
\]
Compared to \eqref{eq:2mTrc2}, we need to change $\xi$ to
$\eps^\alpha \xi_{1}$ or $\eps^\alpha \xi_{-1}$ (the factor
$\eps^\alpha$ comes from the fact that we are looking at the low
frequency regime). Using the notations in the proof of
Lemma~\ref{lem:con2mTr}, we obtain
\begin{equation}
\begin{aligned}
&\lim_{\eps\to 0}\sum_{m,n\geq1}\E\{\eps^{-\alpha d}\FF_{m,\eps}(t,\xi_{1})\FF_{n,\eps}^*(t,\xi_{-1})\}\\
=&\lim_{\eps\to 0}\sum_{\sigma:N_{cr}\geq
  1}\frac{(-1)^n}{(i\eps)^{m+n}}\int_{\sigma_{2k}(t)} dsdu
\int_{\R^{kd}}
dw\prod_{j=1}^{k}\frac{\hat{R}(w_j)}{(2\pi)^d}e^{-\g(w_j)\frac{v_j^+-v_j^-}{\eps^2}}
\\
&\times\prod_{j=1}^{N_{cr}} 
e^{i\mathfrak{s}_j(|\eps^\alpha\xi_{\mathfrak{s}_j}-\ldots-P_{j-1}|^2-|\eps^\alpha\xi_{\mathfrak{s}_j}-\ldots-P_j|^2)\frac{r_j^+-r_j^-}{2\eps^2}}\\
&\times \prod_{j=0}^{N_{cr}}\left(\prod_{l\in \mathcal{A}_j\cup
    \mathcal{B}_j}
e^{i\tau_l(|\eps^\alpha\xi_{\tau_l}-\ldots-P_j|^2-|\eps^\alpha\xi_{\tau_l}-\ldots-P_j-w_l|^2)\frac{v_l^+-v_l^-}{2\eps^2}}\right) 
\prod_{j=1}^{N_{cr}}e^{iP_j\cdot \eta r_j^- \eps^{\alpha+\beta-2}} \\
&\times \frac{1}{\eps^{\alpha
    d}}\hat{\phi}_0(\xi_1-\frac{P_1+\ldots+P_{N_{cr}}}{\eps^\alpha})
\hat{\phi}_0^*(\xi_{-1}-\frac{P_1+\ldots+P_{N_{cr}}}{\eps^\alpha}).
\label{eq:concrco1}
\end{aligned}
\end{equation}
Apart from the change $\xi\mapsto \eps^\alpha \xi_{\pm 1}$, the key
difference between \eqref{eq:concrco1} and \eqref{eq:2mTrc2} is the
extra phase factor
\[
\prod_{j=1}^{N_{cr}}e^{iP_j\cdot \eta r_j^- \eps^{\alpha+\beta-2}}
\]
due to $\eta\neq 0$. Since $\alpha+\beta=2$, this phase factor becomes
\[
\prod_{j=1}^{N_{cr}}e^{iP_j\cdot \eta r_j^-
  \eps^{\alpha+\beta-2}}\mapsto \prod_{j=1}^{N_{cr}}e^{iP_j\cdot \eta
  r_j^-},
\]
and we only need to follow the proof of Lemma~\ref{lem:con2mTr} to
obtain
\begin{equation}
\begin{aligned}
&\lim_{\eps\to 0}\sum_{m,n\geq 1}\E\{\eps^{-\alpha d}\FF_{m,\eps}(t,\xi_{1})\FF_{n,\eps}^*(t,\xi_{-1})\}\\
=&\sum_{N_{cr}=1}^\infty\int_{\Delta_{N_{cr}}(t)} dv\int_{\R^{N_{cr}d}} dP \left(\prod_{j=0}^{N_{cr}}e^{-(v_j-v_{j+1})\red(-P_0-\ldots-P_j)}\right) \left( \prod_{j=1}^{N_{cr}} \red(P_j,-P_0-\ldots-P_{j-1})\right)\\
&\times\delta(-P_1-\ldots-P_{N_{cr}})\prod_{j=1}^{N_{cr}}e^{iP_j\cdot \eta v_j}\left(1_{\alpha\in(0,2)}\|\hat{\phi}_0\|_2^2+1_{\alpha=2}\int_{\R^d}\hat{\phi}_0(\xi+\frac{\eta}{2}-p)\hat{\phi}_0^*(\xi-\frac{\eta}{2}-p)dp\right).
\label{eq:concrco2}
\end{aligned}
\end{equation}
The last factor comes from 
\[
\int_{\R^d}\hat{\phi}_0(\xi+\frac{\eps^\beta\eta}{2}-p)\hat{\phi}_0^*(\xi-\frac{\eps^\beta\eta}{2}-p)dp,
\] 
and the assumption of $\beta=2-\alpha$. 
This finishes the proof.~$\Box$

{\bf  Proof of Lemma~\ref{lem:con4co}.}
The proof is similar to the case when we show the convergence of 
\[
\E\{\U_\eps(t,\xi)^M (\U_\eps^*(t,\xi))^N\}\hbox{ for
  $M,N\in\mathbb{N}$.}
\]
The only difference is that $\xi$ is replaced by
$\xi_1\pm\frac{\eps^\beta \eta_1}{2}$ and
$\xi_2\pm\frac{\eps^\beta \eta_2}{2}$.
First, by following the   proof of~\eqref{eq:corhighto2}, we have
\begin{equation*}
\begin{aligned}
&\lim_{\eps\to0}\E\{\U_\eps(t,\xi_1+\frac{\eps^\beta\eta_1}{2})\U_\eps^*(t,\xi_1-\frac{\eps^\beta\eta_1}{2})\U_\eps^*(t,\xi_2+\frac{\eps^\beta\eta_2}{2})\U_\eps(t,\xi_2-\frac{\eps^\beta\eta_2}{2})\}\\
=&\lim_{\eps\to 0}\E\{\U_\eps(t,\xi_1+\frac{\eps^\beta\eta_1}{2})\U_\eps^*(t,\xi_1-\frac{\eps^\beta\eta_1}{2})\}\E\{\U_\eps^*(t,\xi_2+\frac{\eps^\beta\eta_2}{2})\U_\eps(t,\xi_2-\frac{\eps^\beta\eta_2}{2})\}\\
+&\lim_{\eps\to 0}\E\{\U_\eps(t,\xi_1+\frac{\eps^\beta\eta_1}{2})\U_\eps^*(t,\xi_2+\frac{\eps^\beta\eta_2}{2})\}\E\{\U_\eps^*(t,\xi_1-\frac{\eps^\beta\eta_1}{2})\U_\eps(t,\xi_2-\frac{\eps^\beta\eta_2}{2})\}\\
+&\lim_{\eps\to 0}\E\{\U_\eps(t,\xi_1+\frac{\eps^\beta\eta_1}{2})\U_\eps(t,\xi_2-\frac{\eps^\beta\eta_2}{2})\}\E\{\U_\eps^*(t,\xi_2+\frac{\eps^\beta\eta_2}{2})\U_\eps^*(t,\xi_1-\frac{\eps^\beta\eta_1}{2})\}=I_1+I_2+I_3,
\end{aligned}
\end{equation*}
and to complete the proof, we only need to show $I_2=I_3=0$. 

To study the limit of $I_2$, we take, for example,
\[
\E\{\U_\eps(t,\xi_1+\eps^\beta\eta_1/2)\U_\eps^*(t,\xi_2+\eps^\beta\eta_2/2)\}.
\]
We may follow the proof of Lemma~\ref{lem:concrco} and obtain a phase
factor
\[
\prod_{j=1}^{N_{cr}} e^{iP_j\cdot
  (\xi_1-\xi_2+\eps^\beta(\eta_1-\eta_2)/2)r_j^-\eps^{\alpha-2}},
\]
as in \eqref{eq:concrco1}. Since $\xi_1\neq \xi_2$, the assumption that $\alpha\in (0,2)$
ensures that we have a large phase for multiple scattering; for single
scattering, after change of variable $P_1\mapsto \eps^\alpha P_1$, we
get a factor
\[
e^{i P_1\cdot
  (\xi_1-\xi_2+\eps^\beta(\eta_1-\eta_2)/2)r_j^-\eps^{2\alpha-2}},
\]
so we have a large phase if $\alpha\in (0,1)$. In the end, we only
need to follow the proof of Lemma~\ref{lem:cor2m} to conclude that
$I_2=0$.

For $I_3$, take, for example, 
\[
\E\{\U_\eps(t,\xi_1+\eps^\beta\eta_1/2)\U_\eps(t,\xi_2-\eps^\beta\eta_2/2)\}.
\] 
As in the proof of Lemma~\ref{lem:cor2m}, the corresponding phase
factor becomes 
\[
\prod_{j=1}^{N_{cr}}e^{-i(|P_j|^2-P_j\cdot
(\eps^\alpha\xi_1-\eps^\alpha\xi_2+\eps^{\alpha+\beta}(\eta_1+\eta_2)/2-2(P_0+\ldots+P_{j-1}))r_j^-/\eps^2},
\]
as in \eqref{eq:coreq1}. The rest of discussion is the same, that is
when $\alpha\in (0,1)$, there is always a large phase, which implies
$I_3=0$.  $\Box$

Now we can discuss the limit of $W_\eps$. We use $\F_x,\F_\xi$ to
denote the Fourier transform in $x,\xi$ variable respectively. First,
by the dominated convergence theorem, we have
\begin{equation}
\lim_{\eps\to 0}\E\{\langle W_\eps(t),\varphi\rangle\}=\frac{1}{(2\pi)^d}\int_{\R^{2d}}\lim_{\eps\to 0}\E\{\U_\eps(t,\xi+\frac{\eps^\beta\eta}{2})\U_\eps^*(t,\xi-\frac{\eps^\beta\eta}{2})\}(\F_{x}\varphi)^*(\eta,\xi)d\eta d\xi.
\label{eq:lim1mWig}
\end{equation}
Using Lemma~\ref{lem:concrco}, we need to discuss the following two cases.

\emph{Case 1:} $\alpha+\beta=2,\alpha\in (0,2)$. 
Using \eqref{eq:concrco2}, we integrate $\eta,\xi$ in \eqref{eq:lim1mWig} to obtain
\begin{equation*}
\begin{aligned}
&\lim_{\eps\to 0}\E\{\langle W_\eps(t),\varphi\rangle\}
=\|\hat{\phi}_0\|_2^2\sum_{N_{cr}=1}^\infty\int_{\Delta_{N_{cr}}(t)}
dv\int_{\R^{N_{cr}d}} dP \\
&\times\left(\prod_{j=0}^{N_{cr}}e^{-(v_j-v_{j+1})\red(-P_0-\ldots-P_j)}\right) \left( \prod_{j=1}^{N_{cr}} \red(P_j,-P_0-\ldots-P_{j-1})\right)\\
&\times\delta(-P_1-\ldots-P_{N_{cr}})(\F_{\xi}\varphi)^*(-\sum_{j=1}^{N_{cr}}P_jv_j,0)=\int_{\R^{2d}}\bar{W}_{\delta,s}(t,x,0)\varphi^*(x,\xi)dxd\xi,
\end{aligned}
\end{equation*}
with 
\begin{equation*}
\begin{aligned}
\bar{W}_{\delta,s}(t,x,\xi)
=&\|\hat{\phi}_0\|_2^2\sum_{N_{cr}=1}^\infty\int_{\Delta_{N_{cr}}(t)} dv\int_{\R^{N_{cr}d}} dP \left(\prod_{j=0}^{N_{cr}}e^{-(v_j-v_{j+1})\red(\xi-P_0-\ldots-P_j)}\right)\\
 \times & \left( \prod_{j=1}^{N_{cr}} \red(P_j,\xi-P_0-\ldots-P_{j-1})\right)\delta(\xi-P_1-\ldots-P_{N_{cr}})\delta(x-\xi t+\sum_{j=1}^{N_{cr}}P_jv_j).
\end{aligned}
\end{equation*}
Clearly, we have 
\[
\bar{W}_{\delta,s}(t,x,\xi)=\bar{W}_\delta(t,x,\xi)-\|\hat{\phi}_0\|_2^2\delta(\xi)\delta(x)e^{-\red(0)t},
\]
which consists of the scattering component of the transport equation \eqref{eq:trxxi}
with the initial condition 
\[
\bar{W}_\delta(0,x,\xi)=\|\hat{\phi}_0\|_2^2\delta(\xi)\delta(x).
\] 

\emph{Case 2:} $\alpha=2,\beta=0$. By a similar discussion, we have
\begin{equation}
\begin{aligned}
\lim_{\eps\to0}\E\{\langle W_\eps(t),\varphi\rangle\}=\int_{\R^{2d}}\bar{W}_{\delta,s}(t,x,0)\varphi^*(x,\xi)dxd\xi
\end{aligned}
\end{equation}
with 
\[
\bar{W}_{\delta,s}(t,x,\xi)=\bar{W}_\delta(t,x,\xi)-(2\pi)^d\delta(\xi)|\phi_0(x)|^2e^{-\red(0)t},
\]
and $\bar{W}_\delta(t,x,\xi)$ solving \eqref{eq:trxxi}
with initial condition $\bar{W}_\delta(0,x,\xi)=(2\pi)^d\delta(\xi)|\phi_0(x)|^2$.

By Lemma~\ref{lem:con4co}, if we further assume $\alpha\in (0,1)$, we have
\begin{equation}
\lim_{\eps\to0}\E\{|\langle W_\eps(t),\varphi\rangle|^2\}=|\lim_{\eps\to0}\E\{\langle W_\eps(t),\varphi\rangle\}|^2,
\end{equation}
which implies $\langle W_\eps(t),\varphi\rangle$ converges in probability. 


\appendix

\section{Moments of product of Gaussians}

The following result is standard, we present a proof for the sake of
convenience.  We assume that
\[
\{N_{ij}:i=1,\ldots,m,j=1,\ldots,M_i\}
\]
are zero-mean real (complex)
Gaussian random variables, and write
\begin{equation}
\E\{\prod_{i=1}^m\prod_{j=1}^{M_i} N_{ij}\}=\sum_\F \prod_{((i,j), (\tilde{i},\tilde{j}))\in\F}\E\{N_{ij}N_{\tilde{i}\tilde{j}}\},
\end{equation}
where $\sum_\F$ extends over all pairings formed over vertices
$\{(i,j):i=1,\ldots,m,j=1,\ldots,M_i\}$. 
We set
\[
A_i=\{(i,j):j=1,\ldots,M_i\}.
\]
For a given pairing $\F$ and $i\neq \tilde{i}$, we say that
$A_i$ is connected to
$A_{\tilde{i}}$, and denote this by $A_i\leftrightarrow A_{\tilde{i}}$, if
there exist $j,\tilde{j}$ such that
$((i,j), (\tilde{i},\tilde{j}))\in\F$. In this way, the set
$\{A_i:i=1,\ldots,m\}$ is decomposed into connected components, and we
denote the size of the smallest component by $N_s(\F)$.
\begin{lemma}
For each $i=1,\ldots,m$, let $X_i=\prod_{j=1}^{M_i}N_{ij}$, then we have
  \begin{equation}
    \E\{\prod_{i=1}^m (X_i-\E\{X_i\})\}=\sum_{\F:N_s(\F)\geq 2}\prod_{((i,j), (\tilde{i},\tilde{j}))\in\F}\E\{N_{ij}N_{\tilde{i}\tilde{j}}\}
    \end{equation}
\label{lem:nonself}
\end{lemma}
{\bf Proof.}
We write
\begin{equation}
\begin{aligned}
\E\{\prod_{i=1}^m (X_i-\E\{X_i\})\}=&\E\{X_1\prod_{i=2}^m (X_i-\E\{X_i\})\}-\E\{X_1\}\E\{\prod_{i=2}^m (X_i-\E\{X_i\})\},\\
\end{aligned}
\end{equation}
and note that every term in the expansion of 
\[
X_1\prod_{i=2}^m (X_i-\E\{X_i\})
\]
is a product of zero-mean Gaussians (with possible multiplicative
constant), so when taking expectation, we follow the rule of computing
joint moments of zero-mean Gaussians. For any pairing such that $A_1$
is not connected to any $A_i,i\neq1$, we have a cancellation from the
corresponding term in
\[
\E\{X_1\}\E\{\prod_{i=2}^m (X_i-\E\{X_i\})\}.
\]
Thus, we can write
\begin{equation}
\E\{\prod_{i=1}^m (X_i-\E\{X_i\})\}=\E_1\{X_1\prod_{i=2}^m (X_i-\E\{X_i\})\},
\end{equation}
where $\E_1$ stands for the expectation with the summation over those $\F$ such that
$A_1\leftrightarrow A_i$ for some $i\neq 1$. Following a similar procedure for
$X_2-\E\{X_2\}$, we have
\begin{equation}
\E\{\prod_{i=1}^m (X_i-\E\{X_i\})\}=\E_{1,2}\{X_1X_2\prod_{i=3}^m (X_i-\E\{X_i\})\},
\end{equation}
with $\E_{1,2}$ stands for the expectation with the summation over those $\F$ such that
$A_1\leftrightarrow A_i$ for some $i\neq 1$ and
$A_2\leftrightarrow A_i$ for some $i\neq 2$. In the end, we obtain
\begin{equation}
\E\{\prod_{i=1}^m (X_i-\E\{X_i\})\}=\E_{1,\ldots,m}\{\prod_{i=1}^mX_i\},
\end{equation}
where we only take the expectation with the summation over those $\F$ such that
for all $i=1,\ldots,m$, $A_i\leftrightarrow A_j$ with some~$j\neq i$,
and these are exactly the pairings with $N_s(\F)\geq 2$.~$\Box$



\end{document}